\documentclass{article}
\usepackage{arxiv}

\usepackage[utf8]{inputenc} 
\usepackage[T1]{fontenc}    
\usepackage{hyperref}       
\usepackage{url}            
\usepackage{booktabs}       
\usepackage{tabularx}
\usepackage{amsfonts}       
\usepackage{nicefrac}       
\usepackage{microtype}      
\usepackage{lipsum}		
\usepackage{graphicx}
\usepackage{doi}
\usepackage{subcaption}
\usepackage{forloop}
\usepackage{amsthm}
\usepackage{bbm}
\usepackage{amsmath}
\usepackage{multirow}
\usepackage{soul}
\usepackage{url}
\usepackage{color}
\usepackage{tikz}
\usepackage[numbers]{natbib}

\theoremstyle{remark}
\newtheorem{remark}{Remark}

\theoremstyle{definition}

\newtheorem*{definition}{Definition}

\newcommand{\ptitle}[1]{\textbf{#1}}
\usetikzlibrary{calc}

\newcommand{\ind}[1]{\mathbbm{1}_{\{#1\}}}

\newcommand{\Hcal}{\mathcal{H}}

\newcommand{\Ecal}{\mathcal{E}}
\newcommand{\Ucal}{\mathcal{U}}

\def\eqref#1{equation~\ref{#1}}

\def\1{\bm{1}}

\DeclareMathAlphabet{\mathsfit}{\encodingdefault}{\sfdefault}{m}{sl}
\SetMathAlphabet{\mathsfit}{bold}{\encodingdefault}{\sfdefault}{bx}{n}

\def\Ecal{{\mathcal{E}}}

\def\Hcal{{\mathcal{H}}}

\def\Ucal{{\mathcal{U}}}

\title{Exploring the Performance of Continuous-Time Dynamic Link Prediction Algorithms}

\date{} 					

\author{
	Raphaël Romero\\ 
	Ghent University\\
	\texttt{raphael.romero@ugent.be }
	\And 
	Maarten Buyl \\
	Ghent University\\
	\texttt{maarten.buyl@ugent.be }
	\And 
	Tijl De Bie \\
	Ghent University\\
	\texttt{tijl.debie@ugent.be}
	\And 
	Jefrey Lijffijt \\
	Ghent University\\
	\texttt{jefrey.lijffijt@ugent.be}
}
	
\renewcommand{\headeright}{}

\hypersetup{
pdftitle={A template for the arxiv style},
pdfsubject={q-bio.NC, q-bio.QM},
pdfauthor={David S.~Hippocampus, Elias D.~Striatum},
pdfkeywords={First keyword, Second keyword, More},
}

\begin{document}

\maketitle

\begin{abstract}
	Dynamic Link Prediction (DLP) addresses the prediction of future links in evolving networks. However, accurately portraying the performance of DLP algorithms poses challenges that might impede progress in the field. Importantly, common evaluation pipelines usually calculate ranking or binary classification metrics, where the scores of observed interactions (positives) are compared with those of randomly generated ones (negatives). However, a single metric is not sufficient to fully capture the differences between DLP algorithms, and is prone to overly optimistic performance evaluation. Instead, an in-depth evaluation should reflect performance variations across different nodes, edges, and time segments. In this work, we contribute tools to perform such a comprehensive evaluation. (1) We propose Birth-Death diagrams, a simple but powerful visualization technique that illustrates the effect of time-based train-test splitting on the difficulty of DLP on a given dataset. (2) We describe an exhaustive taxonomy of negative sampling methods that can be used at evaluation time. (3) We carry out an empirical study of the effect of the different negative sampling strategies. Our comparison between heuristics and state-of-the-art memory-based methods on various real-world datasets confirms a strong effect of using different negative sampling strategies on the test Area Under the Curve (AUC). Moreover, we conduct a visual exploration of the prediction, with additional insights on which different types of errors are prominent over time.
\end{abstract}

\def\datasets{
	UNtrade,
	Flights,
	CanParl,
	enron,
	lastfm,
	mooc,
	uci,
	wikipedia}

\section{Introduction}

Many real-world phenomena such as computers networks \cite{yoonFastAccurateAnomaly2019}, epidemics \cite{machensInfectiousDiseaseModel2013}, neural networks \cite{cadenaModelingTemporalNetwork2020}, email exchanges\cite{klimtEnronCorpusNew2004}, and face-to-face interactions \cite{eagleRealityMiningSensing2006,klimtEnronCorpusNew2004} can be modeled as a set of objects interacting through time. This type of data is commonly represented as a dynamic graph \cite{holmeTemporalNetworksModeling2021}, where nodes represent the objects and edges represent pairs of objects that interact through time. While initial attempts at capturing the temporal evolution of networks typically aggregated the interactions into a sequence of static graphs, recent efforts incorporate time continuously, avoiding the loss of some fine-grained temporal information during data preprocessing \cite{holmeTemporalNetworksModeling2021,NaokiMasuda}. The resulting type of data is commonly referred to as Continuous-Time Dynamic Graphs (CTDGs) \cite{poursafaeiBetterEvaluationDynamic2023}.
Modeling and forecasting CTDGs has recently become a very active field of research, as suggested by recent surveys \cite{kazemiRepresentationLearningDynamic2020a,longaGraphNeuralNetworks2023a}.
A crucial task of interest is Dynamic Link Prediction (DLP), where the goal is to predict future links from a history of observed ones.
This task has gained considerable attention, as seen from recent benchmarks \cite{huangTemporalGraphBenchmark2023}, and finds notable applications in recommender systems, influence detection, routing in networks or disease prediction \cite{srinivasApplicationsLinkPrediction2016a,kumarLinkPredictionTechniques2020} to name a few.

Creating a standardized evaluation for Dynamic Link Prediction (DLP) algorithms poses significant challenges \cite{poursafaeiBetterEvaluationDynamic2023}. Firstly, the benchmark datasets available vary widely in nature, leading to different domain-specific DLP tasks. Predicting which pair of students will have a face-to-face interaction in the HighSchool dataset at a given time is for instance very different from predicting which item a given user will interact with in the Wikipedia dataset. Secondly, the evaluation pipelines differ among methods. This often results in near-perfect performance metrics and contributes to a bias where each paper tends to favor its own proposed approach. Lastly, typical metrics for DLP will compare the score of the actual interactions occurring in the Dynamic Graph with the scores of interactions that didn't happen, obtained through random \emph{Negative Sampling (NS)}  \cite{mikolovEfficientEstimationWord2013}. As underlined in \cite{poursafaeiBetterEvaluationDynamic2023}, the procedure used to generate these negative events can have a dramatic impact on DLP performance measures, to the point that some sophisticated methods can often be outperformed by parameter-free heuristics. 

As a result, there is a growing awareness that dynamic link prediction performance measures not only depend on the model quality and the challenging nature of the data but crucially also on the strategy for sampling negative events. Notably, a suggestion proposed by \citet{poursafaeiBetterEvaluationDynamic2023} was to generate more challenging negative samples by examining which edges were previously seen or not at test time. These conclusions align with the guidelines proposed by \citet{junuthulaEvaluatingLinkPrediction2016a}, who suggest splitting the DLP task into two tasks: predicting previously observed links and predicting previously unobserved links, each of these tasks coming with their own specific metrics. This same work attributes these challenges to the fact that, while conventional Machine Learning tasks target independent and identically distributed (iid) data, \emph{events}, \emph{nodes}, and \emph{node pairs} in a (dynamic) graph do not satisfy this property.
As a consequence, the prediction performance is likely to exhibit substantial variations depending on the node, edge, or time interval considered.

Despite this growing awareness, central aspects of the DLP task remain ambiguous to this day. Notably, there is a lack of tools for understanding the domain-dependent effect of splitting a history of interaction into a train history and a test history based on a cutoff time. Yet, such an understanding is crucial for designing relevant NS strategies for evaluation.
Furthermore, the time-evolution of prediction performance tends to be disregarded, despite its significant relevance in real-world applications.

\ptitle{Contributions}
In this work we investigate the open challenges discussed above through visualization and empirical evaluation.
\begin{enumerate}
	\item We introduce the Birth-Death diagram. As illustrated in Fig. \ref{fig:bd_diagram}, this simple plot facilitates the visualization and comparison of the lifetimes of nodes and edges (potentially extending to higher-order structures) in a CTDG. Crucially, this visualization tool enables a clear representation of the partitioning of these objects as influenced by the time-based splitting of the history of events. We discuss two key measures derived from these plots, the \emph{node} and \emph{edge surprise indices}, quantifying the difficulty DLP. We analyze and compare different real-world datasets in terms of these measures.
	\item We demonstrate the utility of Birth-Death diagrams in the design of more useful Negative Sampling (NS) strategies for evaluation. By means of these diagrams, we construct a comprehensive taxonomy that categorizes the types of nodes/edges suitable for use as negative instances against which to contrast the scores of positive events. Subsequently, we leverage this taxonomy to develop more targeted NS strategies specifically intended for evaluation. We assess six key NS strategies derived from this approach and analyze the resulting variations in performance.
	\item Finally, we incorporate time in the evaluation and present a simple visualization method to analyze the time-evolution of Dynamic Link Prediction performance of several recent methods and some heuristics.
\end{enumerate}
Our experiments\footnote{Open-source and free to use code used for the experiments is open source and available at \url{https://github.com/aida-ugent/dlp_exploration.git}} confirm that the performance of methods depend highly on the strategy used for NS. Moreover, the strategy which leads the model to commit more prediction errors (i.e. where the model scores the negative event higher than the positive) varies through time. This observation opens up opportunities for comparing methods empirically by juxtaposing their profile of performance over time.

\ptitle{Outline}
This paper is divided as  follows. In Sec. \ref{sec:related_work} we start by discussing related work on Dynamic Graphs, Dynamic Link Prediction and strategies for evaluating this task. In Sec. \ref{sec:dynamic_graphs} we formally introduce the Birth-Death diagrams, along with corresponding statistics (the node and Edge Surprise indices), which are central to assess the difficulty of the DLP task on a given dataset.  In Sec. \ref{sec:dlp} we subsequently discuss the evaluation of Dynamic Link Prediction algorithms through NS, and propose a taxonomy of the types of negative samples that can be derived. Finally, in Sec. \ref{sec:experiments} and \ref{sec:results} we  conduct numerical experiments to assess the effect of the different NS strategies, and the evolution of the predictive performance over time.


\definecolor{cerulean}{rgb}{0.0, 0.48, 0.65}
\definecolor{deepsaffron}{rgb}{1.0, 0.6, 0.2}
\definecolor{ao(english)}{rgb}{0.0, 0.5, 0.0}

\newcommand{\blue}[1]{\textcolor{cerulean}{#1}}
\newcommand{\orange}[1]{\textcolor{deepsaffron}{#1}}
\newcommand{\green}[1]{\textcolor{ao(english)}{#1}}

\begin{figure}[t]
	\centering
	\includegraphics[width=\textwidth]{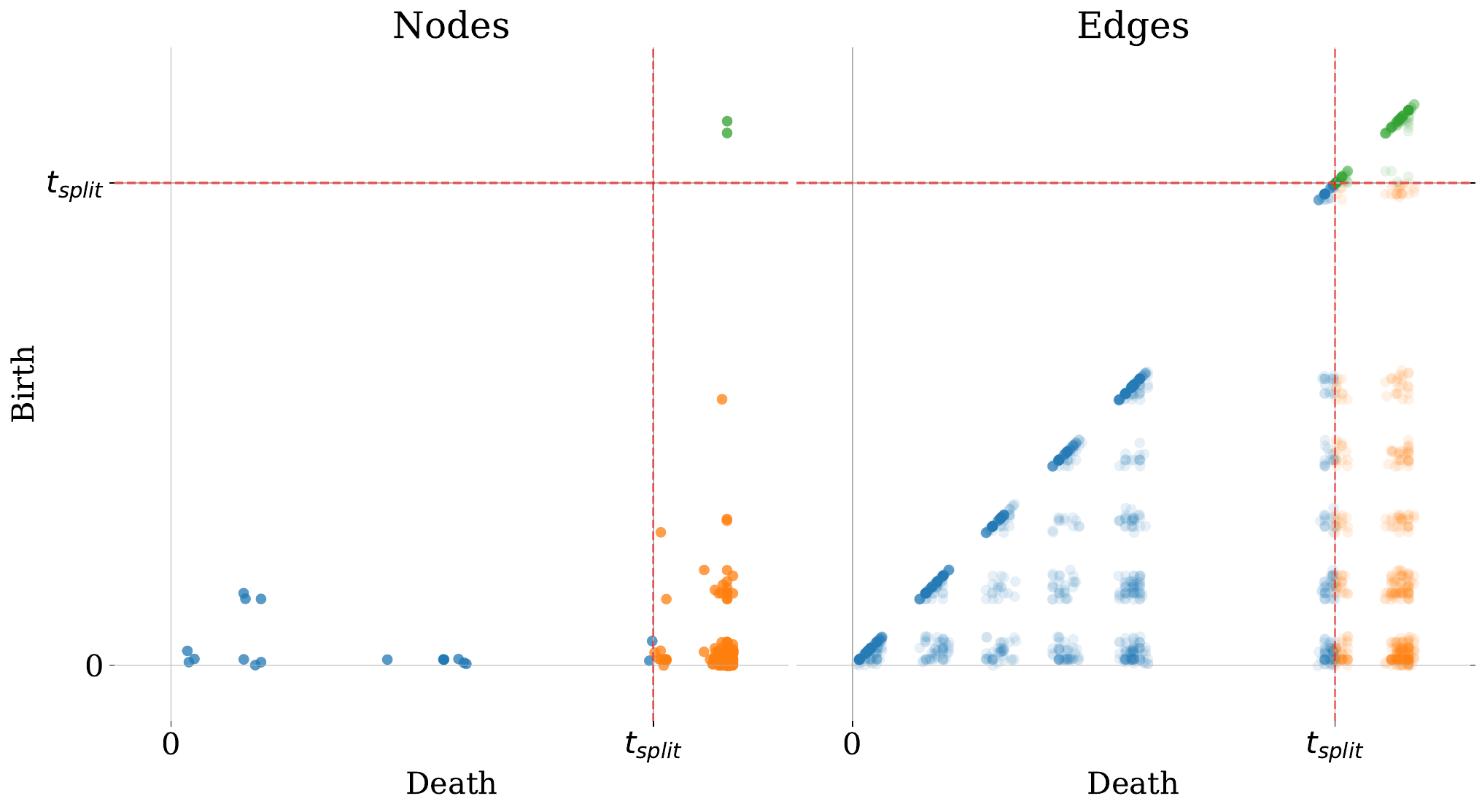}
	\caption{A Birth-Death diagram on a recording of face-to-face interactions between HighSchool students over 9 days \cite{HighSchool_Fournet_Barrat_2014}. The y and x coordinate for each node/edge represent their first (Birth) and last (Death) interaction time respectively. Given a cutoff-time $t_{split}$, while the history of interaction gets divided into a train and a test set, the nodes and edges get partitioned into three categories: \blue{Historical}, \orange{Overlap} and \green{Inductive}. The Surprise Index is the ratio $\frac{\green{Inductive}}{\green{Inductive}+\orange{Overlap}}$.
	}
	\label{fig:bd_diagram}
\end{figure}

\section{Related Work}\label{sec:related_work}

\ptitle{Temporal Networks}. 
Temporal Networks have been utilized to model extensive systems that comprise entities interacting over time, see \citet{NaokiMasuda} and \citet{holmeTemporalNetworks2012a} for a general introduction. As highlighted by \citet{rozenshteinMiningTemporalNetworks2019}, Temporal Networks have been studied under different terminologies, including Dynamic/Temporal Graphs/Networks, depending on the task and the way time is modeled. In particular, some works consider time as discrete \cite{HE201587} and others as continuous \cite{latapyStreamGraphsLink2018}. In the present paper, we consider Continuous-Time Dynamic Graphs, as defined in \cite{poursafaeiBetterEvaluationDynamic2023,Rossi2020}, and the representation used is the "sequence of interaction model of Temporal Networks" \cite{rozenshteinMiningTemporalNetworks2019}. The emphasis is on predicting the events on different time intervals, and not on studying emergent properties. Although our focus is on continuous-time graphs, the methodology proposed in this paper applies to both discrete and continuous-time representations.

\ptitle{Visualizing Temporal Networks.} Due to the additional complexity introduced by the time dimension, visualizing temporal graphs faces unique challenges, as discussed in recent surveys \cite{beckTaxonomySurveyDynamic2017}. As detailed by \citet{linharesDyNetVisSystemVisualization2017a}, the time aspect exacerbates visual clutter issues, which are already common in static networks visualization. They propose Node Activity Maps, a method to visualize node activity over time, but do not consider the evaluation of Dynamic Link Prediction as a use case. Temporal Edge Traffic (TET) and Temporal Edge Activity (TEA) plots proposed by \citet{poursafaeiBetterEvaluationDynamic2023} enable understanding the effect of time-based splitting on the edges. However, as these visualizations focus on visualizing the events directly, edges that are exclusively observed in the training set are represented in the region as those observed both in the train and test set, rendering a visual comparison of these two sets difficult. In contrast, our proposed Birth-Death diagram focuses on directly illustrating the division of nodes and edges into three distinct regions.

\textbf{Methods for DLP.}
(Dynamic) Link Prediction is a longstanding problem, and various classes of methods have been proposed to address it. Early efforts focused on using traditional tools from statistics and network science, often borrowing from existing literature on modeling of static network. These include univariate time-series models \cite{gunesLinkPredictionUsing2016, huangTimeSeriesLinkPrediction2009}, similarity-based methods \cite{liben-nowellLinkpredictionProblemSocial2007}, probabilistic generative models \cite{fouldsDynamicRelationalInfinite,heaukulaniDynamicProbabilisticModels2013,xuStochasticBlockTransition2015,kimNonparametricMultigroupMembership2013}, and matrix and tensor factorization \cite{dunlavyTemporalLinkPrediction2011}.
Nevertheless, with the success of deep learning and representation learning on static graphs, recent approaches have shifted towards using neural networks, as surveyed in \citet{kazemiRepresentationLearningDynamic2020a,longaGraphNeuralNetworks2023a}. Notably, \emph{memory-based} dynamic graph neural networks (DGNNs) such as TGN \cite{Rossi2020} and Dyrep \cite{trivediDyRepLearningRepresentations2018}, use an encoder-decoder architecture. The encoder maps each node to a time-varying representation in a low-dimensional space, while the decoder allows calculating the probability of interactions from the latent representations of the nodes. More generally, the idea of learning a vector representation (embedding), either at the node-level or the edge-level, has been explored in several other methods \cite{congWeReallyNeed2022,luoNeighborhoodawareScalableTemporala,wangInductiveRepresentationLearning2020,wangTCLTransformerbasedDynamic2021,xuInductiveRepresentationLearning2020,yuBetterDynamicGraph2023a}.

It is important to consider that there is currently a lack of fair and objective comparison between these embedding-based techniques and shallow methods such as the ones previously detailed.
Our objective hereby is not to propose a novel method for DLP, but rather to introduce a new performance visualization approach capable of effectively visualizing the DLP task and supporting the performance evaluation of existing methods. Nonetheless, Birth-Death diagrams are particularly pertinent for assessing and diagnosing memory-based DGNNs. Indeed, these models work by maintaining a memory state summarizing the history associated with specific nodes or edges over time. The accuracy and usefulness of this memory state is greatly influenced by the times at which these nodes/edges start (Birth Time) and stop (Death Time) interacting. We emphasize that our study does not consider Negative Sampling for training, which has its own challenges \cite{daniluk2023temporal}, but rather for \emph{evaluation} in Link Prediction.

\ptitle{Challenges in Evaluating DLP algorithms.}
Although many methods for static and dynamic Link Prediction have been proposed in the past, the formal definition and evaluation of this task has been subject to many debates and refinements over the years. Many evaluation methods have been proposed, including set-based metrics \cite{liben-nowellLinkpredictionProblemSocial2007}, Receiver Operator Characteristic (ROC) curves and associate Area Under the ROC Curve (AUC-ROC) \cite{luLinkPredictionComplex2011,lichtenwalterNewPerspectivesMethods2010}, Average Precision \cite{yangEvaluatingLinkPrediction2015}.
While the above studies used the time information mainly for train-test splitting the data into a training and test, \citet{tylendaTimeawareLinkPrediction2009} presented ways to incorporate the time aspect into the method and evaluation, demonstrating its positive impact on performance.
Subsequently, Junuthula et al. \cite{junuthulaEvaluatingLinkPrediction2016a} suggested separating the DLP problem into two tasks: the prediction of either recurring edges or newly observed edges. They propose a metric combining AUC-ROC and Area Under the Precision-Recall Curve to incorporate these two aspects. 

In these studies, the impact of Negative Sampling was often overlooked in the evaluation. More recently, Poursafaei et al. \cite{poursafaeiBetterEvaluationDynamic2023} proposed more challenging negative samples for deep learning-based DLP methods. They introduced three strategies: Random, Historical, and Inductive. In this paper, we extend this literature by proposing a visualization-based method for separating the possible negative samples into categories, with an emphasis on distinctively sampling from Overlap and Historical edges/nodes. Moreover, we introduce a principled way of scrutinizing the changes in performance over time, depending on the negative sampling strategy used for evaluation.

\begin{table}[h!]
	\centering
	\begin{tabular}{|c|p{8cm}|}
		\hline
		\textbf{Notation}                              & \textbf{Description}                                                         \\
		\hline
		$\mathcal{U}$                                  & Set of nodes                                                                 \\
		$T$                                            & Maximal time                                                                 \\
		$\Hcal$                                        & Stream of events $(u, v, t)$                                                 \\
		$(u, v, t)$                                    & Event representing an interaction of node $u$ with node $v$ at timestamp $t$ \\
		$\Hcal_t$                                      & Set of all events in $\Hcal$ that occur up to time $t$                       \\
		$\Ucal_t$                                      & Set of nodes involved in interactions up to time $t$                         \\
		$\Ecal_t$                                      & Set of edges involved in interactions up to time $t$                         \\
		$\Hcal^u$                                      & Set of events in $\Hcal$ that involve node $u$                               \\
		$\Hcal^{(u,v)}$                                & Set of events in $\Hcal$ that involve edge $(u,v)$                           \\
		$t_{split}$                                    & Cutoff time for train-test split                                             \\
		$\Hcal_{train}$                                & Train set of events occurring up to time $t_{split}$                         \\
		$\Hcal_{test}$                                 & Test set of events occurring after time $t_{split}$                          \\
		$b_\Hcal^x$                                    & Birth time of node or edge $x$ in $\Hcal$                                    \\
		$d_\Hcal^x$                                    & Death time of node or edge $x$ in $\Hcal$                                    \\
		\hline
	\end{tabular}
	\caption{Notation used throughout the paper.}
\end{table}

\section{Understanding the Effect of Splitting a Dynamic Graph based on Time\label{sec:dynamic_graphs}}

In this section, we provide some background on Continuous-Time Dynamic Graphs (CTDGs). Subsequently, we introduce the notions of birth and death time, and the associated Birth-Death diagrams, a visualization tool that allows one to understand the effect of splitting a dynamic graph based on time. Based on this tool, we introduce the node and Edge Surprise indices as metrics for quantifying the difficulty of predicting future links on a given dynamic graph dataset. 


\subsection{Background: Continuous-Time Dynamic Graphs}

For a set of nodes $\mathcal{U}$ and maximal time $T$, a \emph{Continuous-Time Dynamic Graph} (CTDG) is defined through a stream $\Hcal = \{(u, v, t)\} \subset \mathcal{U} \times \mathcal{U} \times [0, T]$ of \emph{events} $(u, v, t)$ that each represent an interaction of the \emph{source} node $u$ with the \emph{destination} node $v$ at timestamp $t$. We use the term \emph{edge} to refer to a pair of nodes $(u, v)$ at an unspecified time. In \emph{directed} graphs, such an edge $(u,v)$ in an event $(u,v,t)$ is an ordered pair of nodes; in other words this means that node $u$ sends an interaction to $v$ at time $t$. Conversely, \emph{undirected} graphs treat edges as unordered pairs of nodes $\{u,v\}$; so an interaction $(u,v,t)$ means that $u$ and $v$ interacted at time $t$. In practice, an undirected edge can be uniquely identified by $(\min(u,v),\max(u,v))$. Further, note that CTDGs allow events to occur at any continuous-valued timestamp $0 \leq t \leq T$ and allow multiple events to happen at the same time.
\newcommand{\triangleq}{\stackrel{\triangle}{=}}

To index the collection of all events $\Hcal$ in the CTDG, we also introduce some helpful shorthand notations. 
We use $\Hcal^{u} \triangleq \{(u', v', t) \in \Hcal \mid u'=u \vee v'=u\}$ to represent the set of all events in $\Hcal$ that involve the \emph{node} $u$ and $\Hcal^{(u,v)} \triangleq \{(u', v', t) \in \Hcal \mid u'=u \wedge v'=v\}$ to represent the set of events that involve the \emph{edge} $(u, v)$. We also define $\Hcal_t$ as the subset of all events in $\Hcal$ that occur \emph{up to} a certain time $t$, i.e. $\Hcal_t \triangleq \{(u, v, t') \in \Hcal \mid t' < t\}$.

Overall, our goal is to better inform evaluation of Dynamic Link Prediction over CTDGs. We hold off on formally introducing this task until Sec.~\ref{sec:dlp} and first consider a core decision in any machine learning evaluation: how the data is split up into training and testing data. A typical assumption in dynamic graphs is that we will only need to make predictions about future events. Hence, the train-test split is commonly determined by a \emph{cutoff time} $t_{\text{split}}$ that partitions the set of events $\Hcal$ into the train set of past, known events $\Hcal_{\text{train}} = \Hcal_{t_{\text{split}}}$ and the test set of `future', unknown events $\Hcal_{\text{test}} = \Hcal \setminus \Hcal_{t_{\text{split}}}$.
In what remains of this section, we characterize nodes and edges by whether they are active exclusively in the train set, test set, or in both.

\subsection{The Birth and Death of Nodes and Edges}\label{sec:bd}
Dynamic graphs dynamically evolve over time. A key motivation for our contributions is that many real CTDG datasets only have nodes and edges that interact \emph{within a specific timeframe}. For instance, in a social network, a new user may join (represented as a node), or a pair of users (i.e. an edge) may cease interacting entirely at a certain time.
To formalize these concepts, we introduce the following definitions:

\begin{definition}[Birth Time]

	For any node $x=u$ or edge $x=(u,v)$, the \textbf{birth time} $b_\Hcal^{x}$ is defined as the earliest time at which this node or edge is involved in an event in the history $\Hcal$:
	\begin{equation}
		b_\Hcal^{x} \triangleq \min\limits_{(u, v, t) \in \Hcal^{x}} t.
	\end{equation}
\end{definition}

\begin{definition}[Death Time]
	For any node $x=u$ or edge $x=(u,v)$, the \textbf{death time} $d_\Hcal^{x}$ is defined as the latest time at which this node or edge is involved in an event in the history $\Hcal$:
	\begin{equation}
		d_\Hcal^{x} \triangleq \max\limits_{(u, v, t) \in \Hcal^{x}} t.
	\end{equation}
\end{definition}

Note that, by definition, $b_\Hcal^{x} \leq d_\Hcal^{x}$.
These definitions allow us to capture the lifespan of nodes and edges in dynamic graphs, which is crucial for understanding their behavior and evolution over time.

We argue that the birth and death times of nodes and edges are highly relevant when splitting up a CTDG's events into a train set $\Hcal_{\text{train}}$ and test set $\Hcal_{\text{test}}$. Such splits are done to assess a model's ability to generalize to unseen data that is encountered in real-world applications, but CTDGs typically see nodes and edges \emph{reoccur} often. In fact, exploiting recurring patterns is an implicit goal of any machine learning task. Previous work \cite{junuthulaEvaluatingLinkPrediction2016a,poursafaeiBetterEvaluationDynamic2023} has hypothesized that it is far easier for any parametrized model to predict if and when an edge occurs in the test set $\Hcal_{\text{test}}$, if it has already learned from the occurrences of the same edge in the train set $\Hcal_{\text{train}}$.
The formal definition of the birth and death times helps to elucidate this assumption. For instance, we can state that the occurrence of an edge in the test set will seem more likely to a model if its birth time $b_\Hcal^{x}$ was before $t_{\text{split}}$. Likewise, nodes that were already active in the train set, i.e. they were `born' at time $b_\Hcal^{x} < t_{\text{split}}$, will be better understood and less surprising than nodes with a birth time $b_\Hcal^{x} \geq t_{\text{split}}$.

Moreover, the extent to which different methods are capable of accurately predicting previously unseen edges in the test set may vary between these different situations. Understanding such differences may be important for choosing the most appropriate method in a particular application.

Therefore, a prudent and useful analysis of (predictions over) a CTDG benefits from partitioning nodes and edges into three categories, which we define here.
In all definitions, we denote by $t_{split}$ the time at which the train-test split is made.

\begin{definition}[Historical]
	A \textbf{Historical} (H) node or edge $x$ only occurs in the train set $\Hcal_{\text{train}}$ and never in the test set $\Hcal_{\text{test}}$, i.e.
	\begin{equation}
		d_\Hcal^{x} < t_{\text{split}}.
	\end{equation}
\end{definition}

\begin{definition}[Inductive]
	An \textbf{Inductive} (I) node or edge $x$ \emph{only} occurs in the test set $\Hcal_{\text{test}}$ and never in the train set $\Hcal_{\text{train}}$, i.e.
	\begin{equation}
		b_\Hcal^{x} \geq t_{\text{split}}.
	\end{equation}
\end{definition}

\begin{definition}[Overlap]
	An \textbf{Overlap} (O) node or edge $x$ occurs in both the train set $\Hcal_{\text{train}}$ \emph{and} the test set $\Hcal_{\text{test}}$, i.e.
	\begin{equation}
		b_\Hcal^{x} < t_{\text{split}} \land d_\Hcal^{x} \geq t_{\text{split}}.
	\end{equation}
\end{definition}




\subsection{The Birth-Death Diagram}\label{sec:bd_diagram}

\begin{figure}
		\begin{tabular}{cc}
			\begin{subfigure}[t]{0.47\linewidth}
				\centering
				\includegraphics[width=\linewidth]{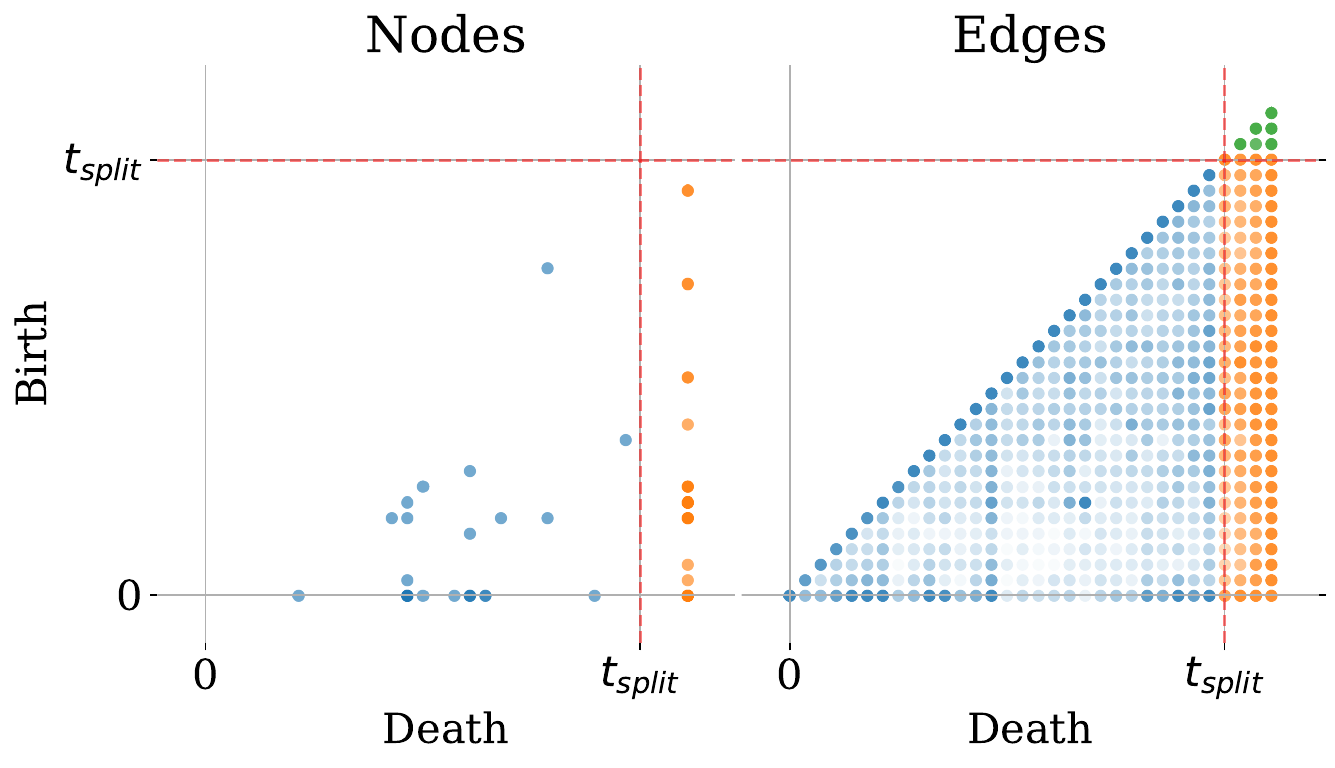}
				\caption{UNtrade}
			\end{subfigure}&
			\begin{subfigure}[t]{0.47\linewidth}
				\centering
				\includegraphics[width=\linewidth]{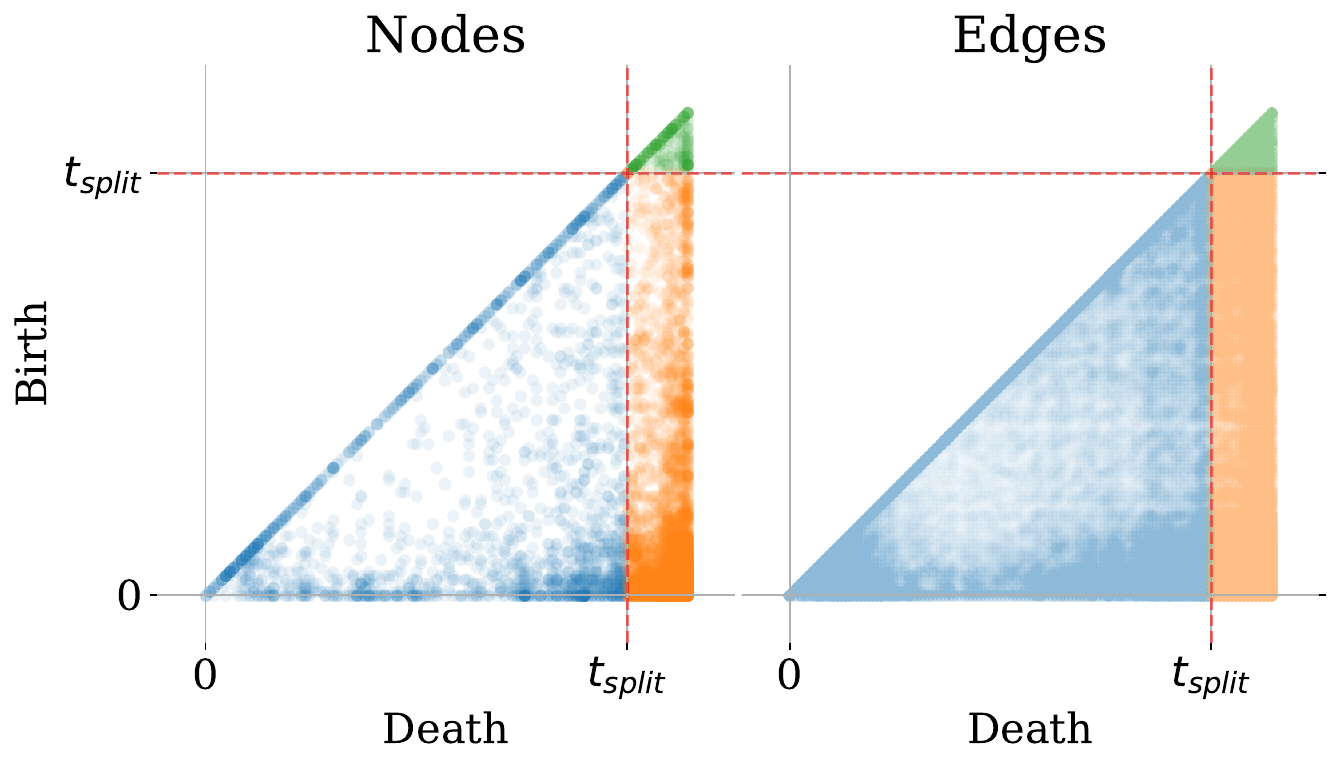}
				\caption{Flights}
			\end{subfigure}\\
			\begin{subfigure}[t]{0.47\linewidth}
				\centering
				\includegraphics[width=\linewidth,  trim={0 0 0 1cm} , clip ]{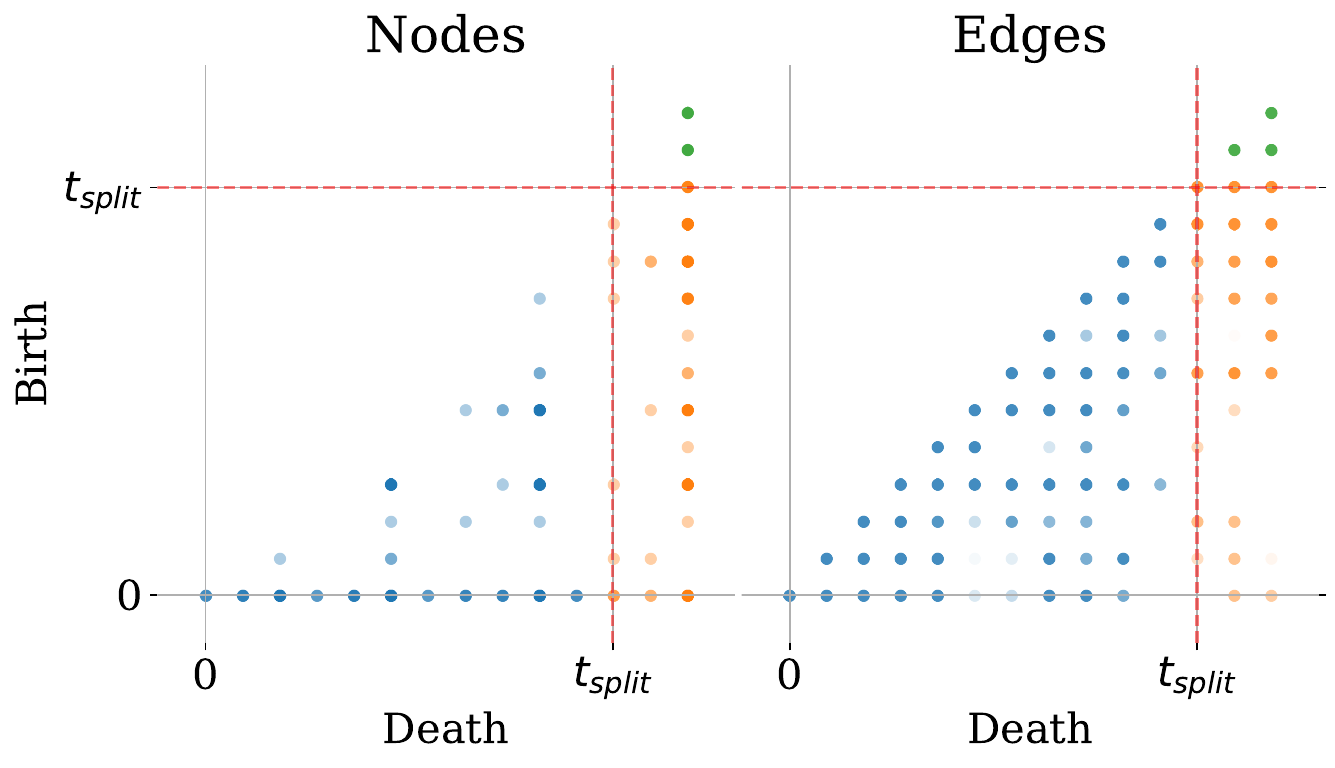}
				\caption{CanParl}
			\end{subfigure}&
			\begin{subfigure}[t]{0.47\linewidth}
				\centering
				\includegraphics[width=\linewidth,trim={0 0 0 1.5cm}, clip]{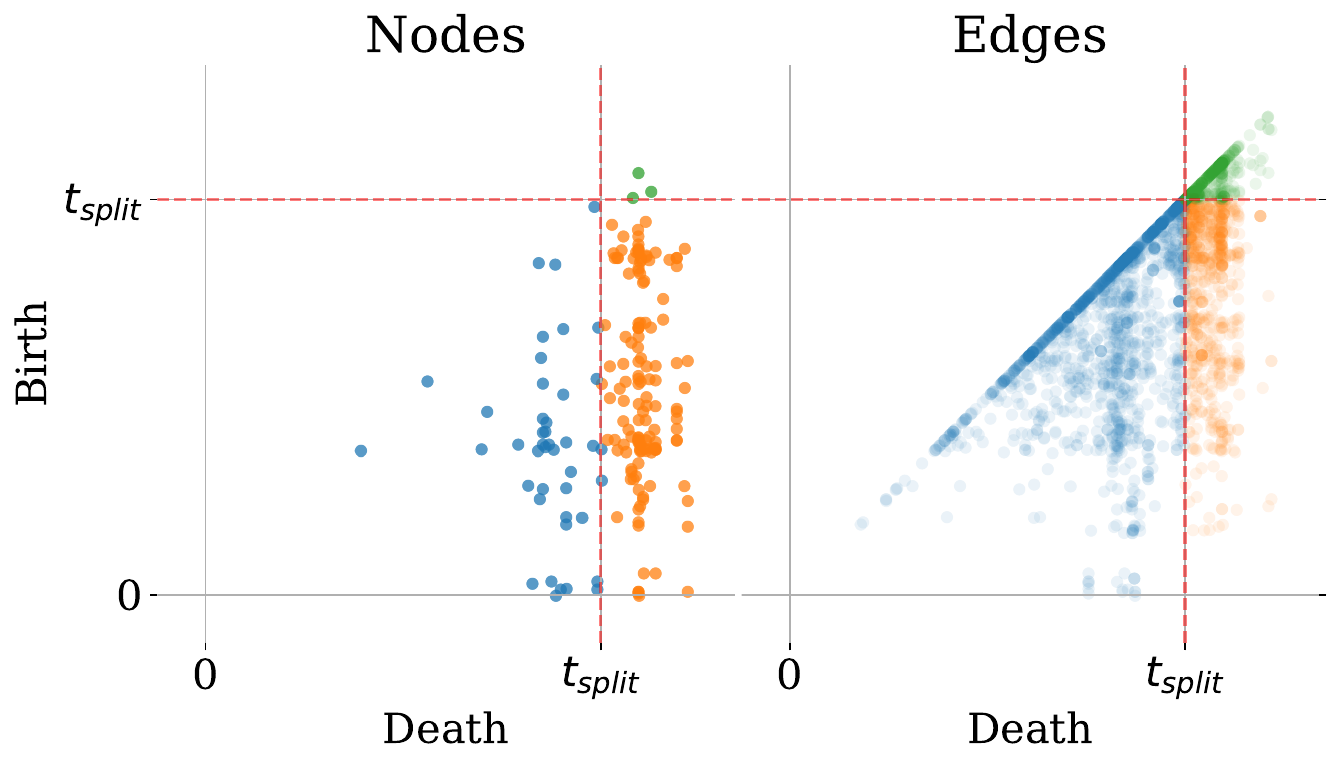}
				\caption{enron}
			\end{subfigure}\\
			\begin{subfigure}[t]{0.47\linewidth}
				\centering
				\includegraphics[width=\linewidth,trim={0 0 0 1.5cm}, clip]{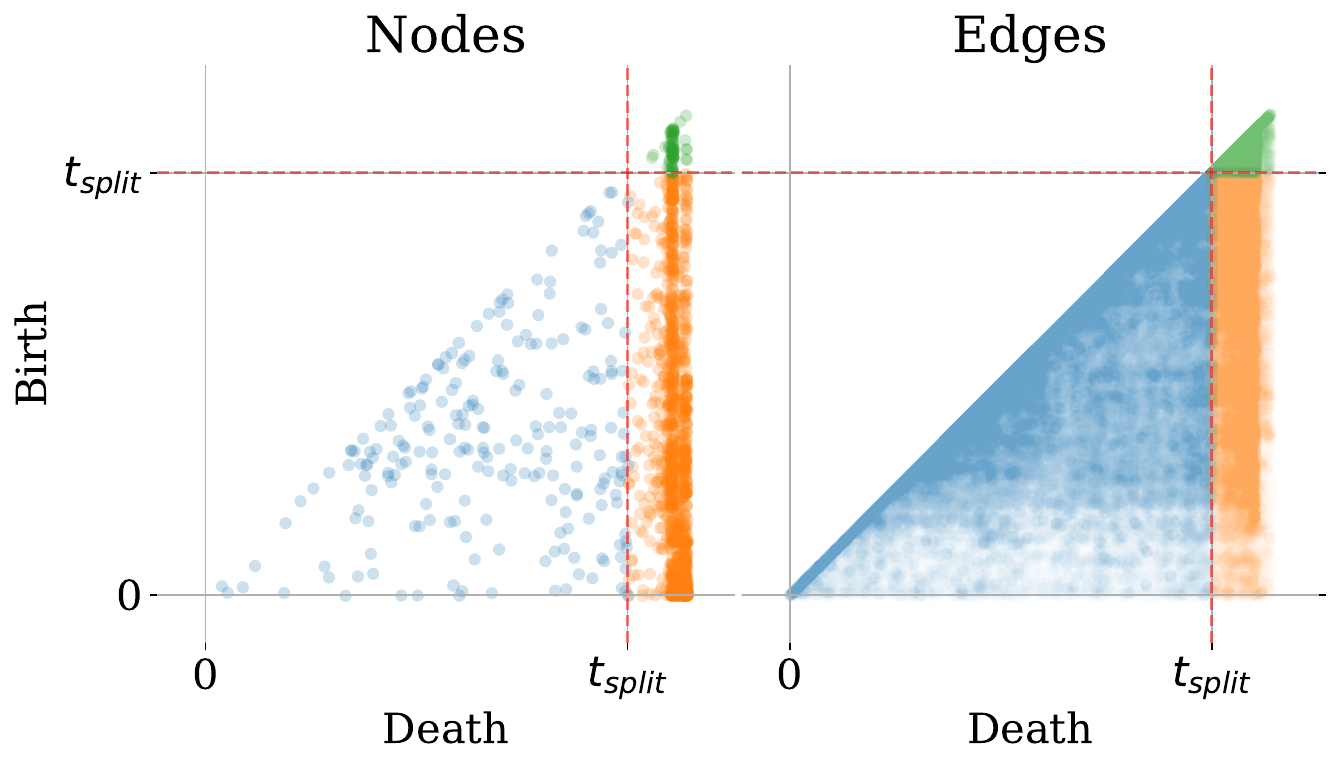}
				\caption{lastfm}
			\end{subfigure}&
			\begin{subfigure}[t]{0.47\linewidth}
				\centering
				\includegraphics[width=\linewidth,trim={0 0 0 1.5cm}, clip]{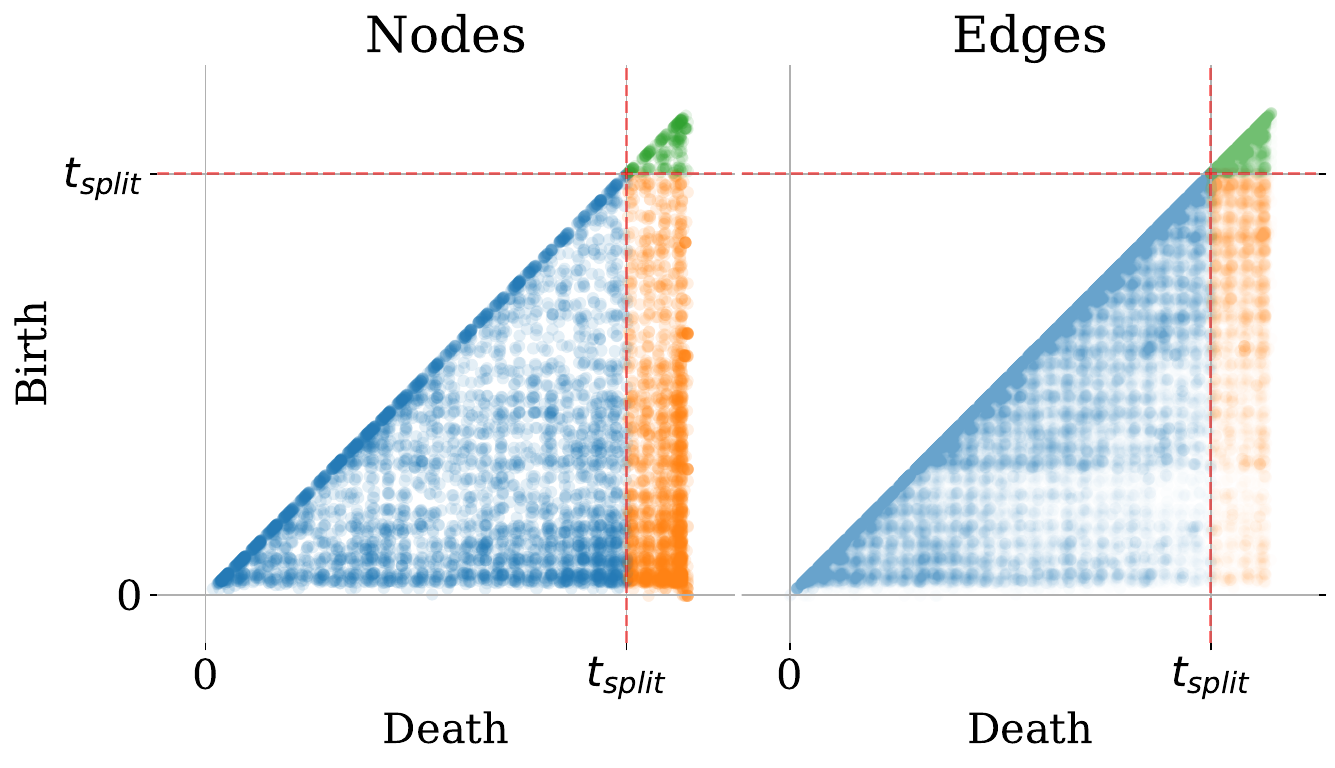}
				\caption{mooc}
			\end{subfigure}\\
			\begin{subfigure}[t]{0.47\linewidth}
				\centering
				\includegraphics[width=\linewidth,trim={0 0 0 1.5cm}, clip]{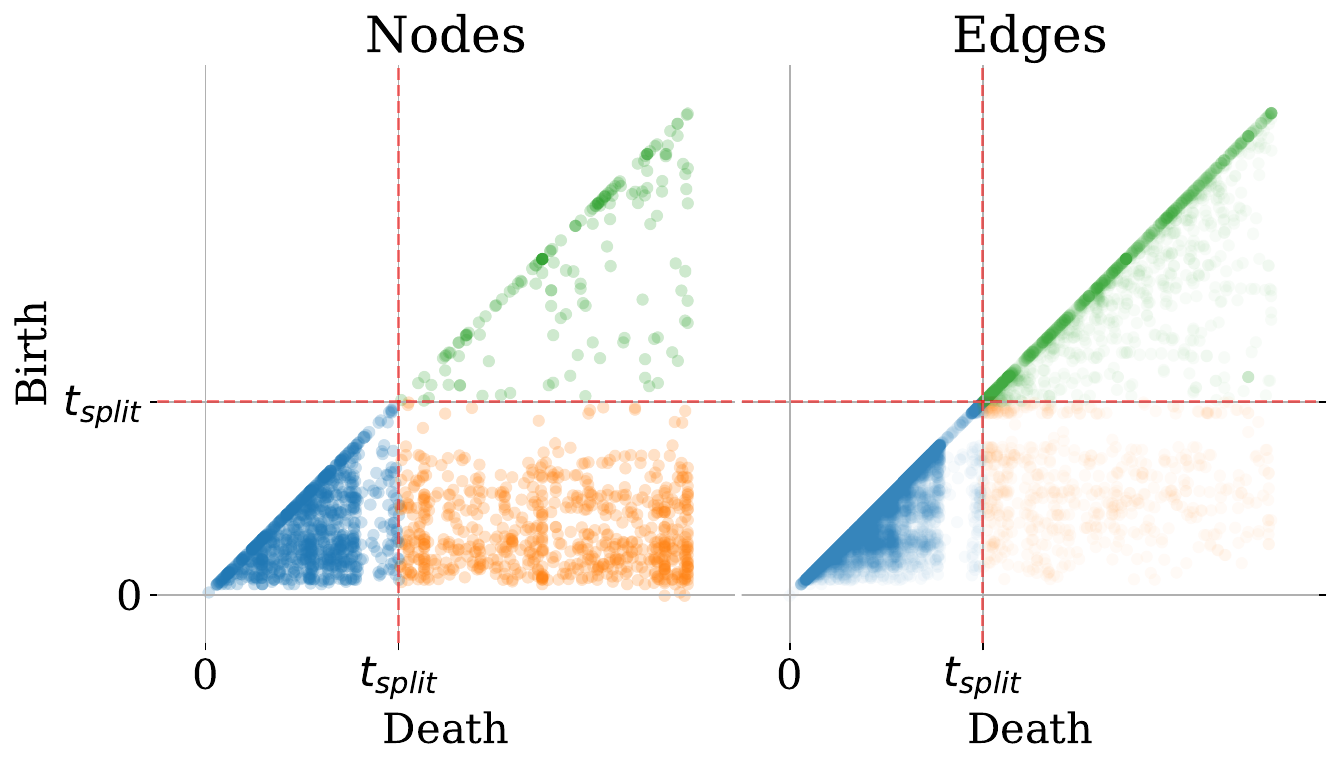}
				\caption{uci}
				\label{fig:bd_uci}
			\end{subfigure}&
			\begin{subfigure}[t]{0.47\linewidth}
				\centering
				\includegraphics[width=\linewidth,trim={0 0 0 1.5cm}, clip]{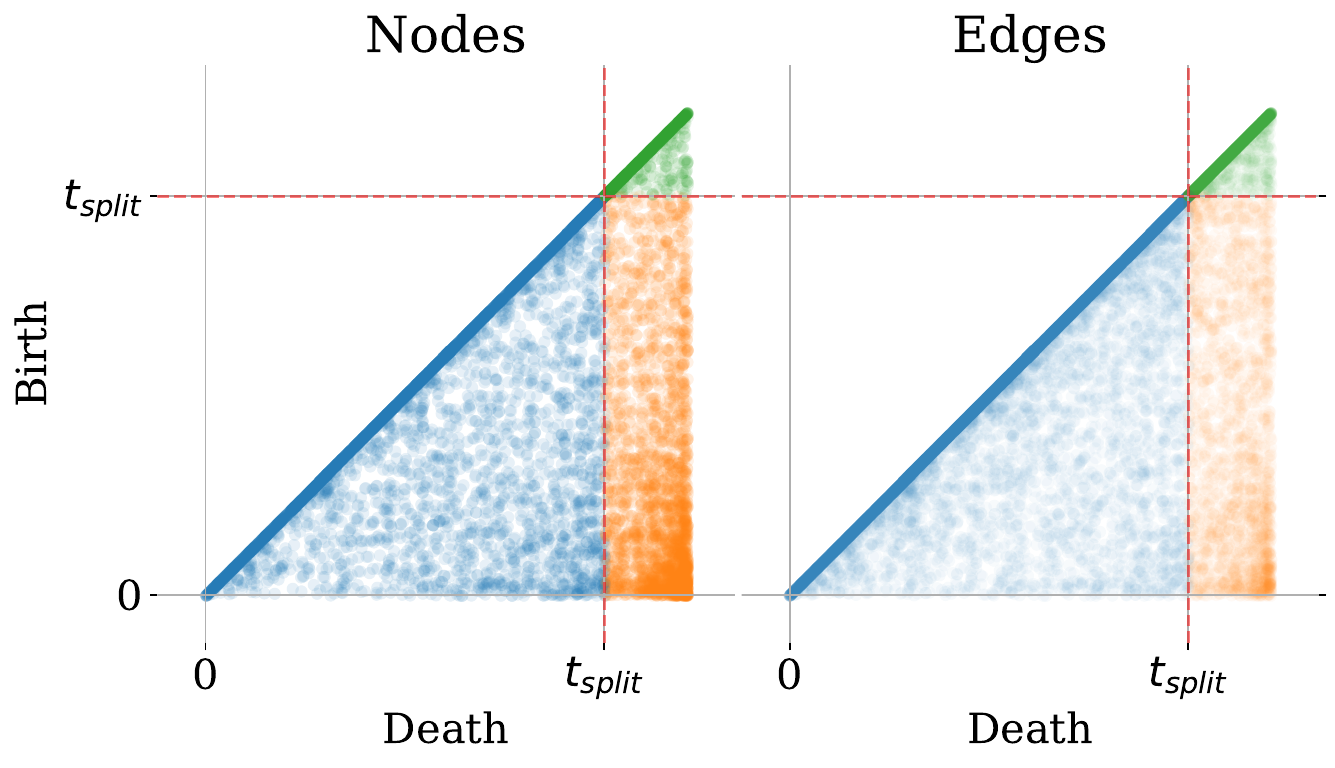}
				\caption{wikipedia}
			\end{subfigure}\\
		\end{tabular}
		
	\caption{Birth-Death diagrams for Nodes and Edges in datasets from the Dynamic Graph Benchmark from \citet{poursafaeiBetterEvaluationDynamic2023}. The datasets are split into train and test sets containing 85\% and 15\% of the events, respectively.
		\label{fig:bd_examples_edges}
	}
\end{figure}
For a given a cutoff time $t_{\text{split}}$, the birth and death times of nodes and edges clearly distinguish whether they are historical, Overlap, or Inductive. By extension, the distribution of their birth and death times characterizes the difficulty of the test set for any cutoff time.

We therefore introduce the \textit{Birth-Death diagram}: a scatter plot visualization that represents each node or edge by its birth time (on the y-axis) and death time (on the x-axis). Fig.~ \ref{fig:bd_diagram} illustrates the Birth-Death diagram on a dataset of Face-to-Face interactions between High School Students. Moreover, in Fig.~\ref{fig:bd_examples_edges}, we plot the Birth-Death diagrams for different CTDG datasets\footnote{More information about the datasets is provided in the Appendix.} from the benchmark of \citet{poursafaeiBetterEvaluationDynamic2023}.

\begin{remark}
	For all datasets in our illustration, the cut-off time is determined as the $1-\alpha$-th quantile of the event times, where $\alpha$ is a train-test split ratio set to $\alpha=0.15$.
	In simpler terms, this means that the cut-off time $t_{split}$ is set to the point in time beyond which  15\% of the events occur. Any events that occur before this point in time are included in the training set, while any events that occur after this point in time are included in the test set.
\end{remark}

From Fig. \ref{fig:bd_examples_edges} we can draw some interesting observations, which we discuss here. 

\ptitle{Seasonality of birth and death times.} The \textit{seasonality} of lifespan patterns can be observed in both the High School and MOOC datasets as the points corresponding to edges and nodes tend to cluster into squares representing days. The UCI dataset, which describes online interactions between students from April to October 2004, also shows seasonality in the holiday break: there is a white stripe during the holiday break (slightly before $t_{split}$). This seasonality is a crucial property of the Link Prediction task at hand. In the case of the HighSchool dataset, it means that we are observing a few complete days of interactions between the students, and that we are trying to predict when and who will interact in the following days. For these datasets, carefully representing time using techniques such as time encoding \cite{xuInductiveRepresentationLearning2020} can be crucial to get good performances.

\ptitle{Short-lived nodes and edges.} The high density of points on the diagonal of these diagrams, in particular in the Wikipedia dataset, indicates that most nodes and edges have very \textit{short lifespans}.
As a result, information learned by models about these instances has a higher chance of becoming obsolete after some time. There, it is crucial that methods take into account time and prioritize attend more to active edges/nodes.
Conversely, however, some nodes and edges have very long lifetimes in the Wikipedia and Flights datasets (their scatter points are in the lower right), so memorization may work well on these.

\ptitle{"Easy" datasets.} There are some datasets, namely UNtrade, Enron, CanParl, and HighSchool, where most of the nodes are observed at least once in the train set, with only a few nodes starting interactions in the test set. Memorization heuristics such as Preferential Attachment and EdgeBank will constitute be strong baselines for these datasets.

Finally, it is crucial to note that for User-Item graphs such as wikipedia, mooc or lastfm, the Birth-Death diagram will yield different profiles for the Users and Item nodes. We showcase and discuss these differences in Appendix \ref{appendix:user_item}.

\subsection{The Surprise Index\label{sec:surprise}}

\begin{figure}[t]
	\centering
	\includegraphics[width=0.7\linewidth]{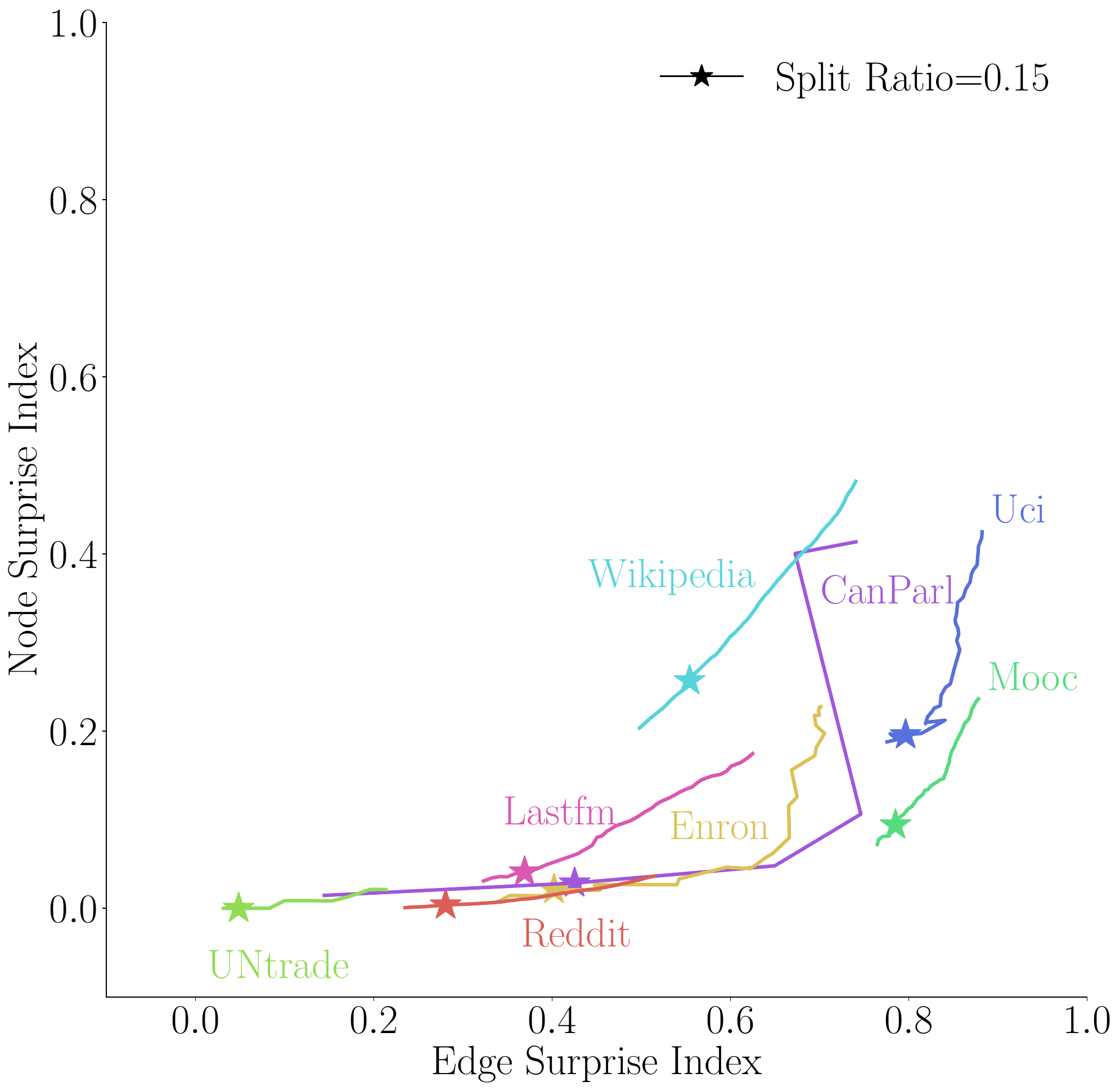}
	\caption{
		Changing the test-split ratio linearly from 0.1 to 0.5 changes the node and Edge Surprise Index differently depending on the dataset. The typical test-split ratio of 0.15 is marked as a "*" on the lines.
	}
	\label{fig:surprise_index}
\end{figure}


The Birth-Death diagrams suggest that the proportion of Overlap and Inductive nodes and edges is highly dependent on the cutoff time $t_{\text{split}}$. We formally assess this proportion through the \emph{Surprise Index}, i.e. the proportion of inductive nodes/edges in the test set.
\begin{definition}
	The Node/Edge \textbf{Surprise Index} is defined as the proportion of nodes and edges $x$ in the test set (i.e. their death time $d_\Hcal^{x} \geq t_{\text{split}}$) that \textit{only} appear in the test set (i.e. their birth time $b_\Hcal^{x} > t_{\text{split}}$). In other words, it is the ratio between the number of Inductive (i.e. having a birth time after $t_{split}$) and the number of Inductive or Overlap (i.e. having a death time after $t_{split}$) nodes/edges. Mathematically, considering $x$ to be either the nodes or edges:
	\begin{align*}
		surprise^{x}=
		 \frac{
			\left|\{x|b_\Hcal^{x} \geq t_{\text{split}}\}\right|
		}{
			\left|\{x|d_\Hcal^{x} \geq t_{\text{split}}\}\right|
		}.                                                              \\
	\end{align*}
	Indeed, by definition the Inductive nodes/edges are those whose birth time is after $t_{split}$, while objects which are either Inductive or Overlap are those whose \emph{death time} is after $t_{split}$.

\end{definition}
\newcommand{\nodesurprise}{$
		\frac{|\Ucal_{test}\setminus \Ucal_{train}|}{|\Ucal_{test}|}
	$}
\newcommand{\edgesurprise}{$
		\frac{|\Ecal_{test}\setminus \Ecal_{train}|}{|\Ecal_{test}|}
	$}

Typical ML pipelines will learn from the training data by hinting on recurring patterns (concepts present in the data) and learn to recognize these patterns. The test data may reproduce these patterns to some extent, along with some other signals previously unobserved by the model, which will be reflected in the fact that the predictions will not be imperfect. In the case of the DLP, we dispose of concrete ways of measuring the quantity of information in the test set that will be new for the model. The Surprise Index is a natural way of measuring that.

In Figure \ref{fig:surprise_index}, we present the Node Surprise Index against the Edge Surprise Index for various datasets, with train-test splitting ratios ranging from 0.1 to 0.5. By examining these indices, we can observe some interesting properties.

\ptitle{The Surprise Index is not necessarily monotonous with the size of the test set.} It may be intuitive to assume that the Surprise Index increases monotonically with a larger proportion of events included in the test set (i.e., earlier cutoff times $t_{\text{split}}$). However, this is not always the case. For instance, in the CanParl dataset, increasing the test ratio from 0.3 to 0.4 actually decreases the Edge Surprise Index, while substantially increasing the Node Surprise Index. A similar non-monotonicity can be observed for the Enron dataset, where increasing the test ratio from 0.3 to 0.4 similarly decreases the Edge Surprise Index. This paradox arises from the fact that the Surprise Index is a non-decreasing function of the ratio between Inductive events (those that occur only in the test set) and Overlapping events (those that occur in both the training and test sets). Thus, if adding more events to the test set increases the number of Overlapping events faster than it increases the number of Inductive events, the Surprise Index will actually decrease. This observation highlights the \emph{importance of carefully selecting cutoff times} to ensure that the evaluation setting accurately reflects the difficulty and type of task at hand.

\ptitle{The Edge Surprise Index is typically Higher than the Node Surprise.} All datasets evaluated here exhibit curves in the lower right of Figure \ref{fig:bd_examples_edges}, indicating that the Edge Surprise Index is generally higher than the Node Surprise Index. While this may seem obvious, we stress here that it doesn't have to be the case. Indeed, what is clear is that if a given node was never observed during training but starts to interact during testing, then it means that the corresponding \emph{edge} it forms during testing was necessarily never seen during training.
As a consequence there are always at least as many Inductive edges as there are Inductive nodes. However, the Surprise Index increases with the \emph{ratio} between the number of Inductive and Overlap nodes/edges, and this ratio doesn't necessarily have to be larger for edges than for nodes.
For instance, suppose that four nodes A, B, C and D interact with each other during both the training and test period, resulting in a total of 6 Overlap edges. Now, suppose that a node E starts interacting with A during testing. In this case the number of Inductive nodes and edges are both 1. However, the Node Surprise Index is $\frac{1}{5}$, while the Edge Surprise Index is $\frac{1}{7}$, which is smaller. In contrast, if E had started interacting with all four nodes, then we would have an Edge Surprise of $\frac{4}{10}$ which would be larger than the Node Surprise.
Another example would be if two nodes E and F would start interacting but only with each other during testing. In this case, the Node Surprise Index would be $\frac{2}{6}$, while the Edge Surprise Index would be $\frac{2}{8}$ which is smaller.
As such it is in itself an interesting pattern that the Edge Surprise Index is generally higher than the Node Surprise Index in all the considered datasets. Reporting both these indices in practice may give a good first overview of the difficulty of the DLP task at hand.

\ptitle{Domain Dependency.} The difference in growth rates between the Node and Edge Surprise indices varies widely across different datasets. For some datasets, increasing the number of events in the test set will increase the proportion of nodes present in the test set that were not observed in the training set. For instance, in the CanParl dataset, increasing the test ratio from 0.1 to 0.3 increases the Edge Surprise Index from 0.15 to around 0.78, while the Node Surprise Index only undergoes a 0.1 increase. This means that while in both cases most of the nodes will already have been observed in the training set, the number of previously unobserved edges will increase significantly. This illustrates the fact that seemingly small changes in the evaluation setting can have a dramatic impact on the difficulty and type of task at hand.
The exact values of the node and Edge Surprise indices, on the datasets from \citet{poursafaeiBetterEvaluationDynamic2023}, are provided in Table \ref{tab:dataset_stats}.

\begin{table}
	\caption{Dataset Statistics, including the Node and Edge Surprise Index, for a test-ratio of 15\%. For most datasets the nodes are mostly all observed during training, hence a relatively low Node Surprise Index. The Edge Surprise is higher however, since many edges that interact in the test set were never observed in the test set.
	}
	\label{tab:dataset_stats}
		\centering
		\resizebox{0.85\pdfpagewidth}{!}{
			\begin{tabular}{l|r|rrrrc|ccccc} 
				\toprule\toprule
						   & Events  & \multicolumn{5}{|c|}{Nodes}                           & \multicolumn{5}{c}{Edges}                            \\ 
				\hline
						   &         & Total & Historical & Overlap & Inductive & Surprise  & Total  & Historical & Overlap & Inductive & Surprise  \\ 
				\midrule
				uci        & 59835   & 1899  & 1052       & 681     & 166       & 0.196     & 20296  & 17069      & 657     & 2570      & 0.796     \\
				highschool & 45047   & 180   & 18         & 160     & 2         & 0.012     & 2239   & 1577       & 445     & 217       & 0.328     \\
				wikipedia  & 157474  & 9227  & 5663       & 2648    & 916       & 0.257     & 18257  & 13667      & 2046    & 2544      & 0.554     \\
				enron      & 125235  & 184   & 43         & 138     & 3         & 0.021     & 3125   & 1914       & 724     & 487       & 0.402     \\
				USLegis    & 60396   & 225   & 112        & 101     & 12        & 0.106     & 26423  & 18884      & 5857    & 1682      & 0.223     \\
				UNvote     & 1035742 & 201   & 7          & 194     & 0         & 0.000     & 31516  & 4487       & 26583   & 446       & 0.017     \\
				UNtrade    & 507497  & 255   & 27         & 228     & 0         & 0.000     & 36182  & 10595      & 24347   & 1240      & 0.048     \\
				SocialEvo  & 2099519 & 74    & 12         & 62      & 0         & 0.000     & 4486   & 2609       & 1827    & 50        & 0.027     \\
				mooc       & 411749  & 7144  & 4732       & 2185    & 227       & 0.094     & 178443 & 147612     & 6628    & 24203     & 0.785     \\
				Flights    & 1927145 & 13169 & 2182       & 10698   & 289       & 0.026     & 395072 & 264189     & 85369   & 45514     & 0.348     \\
				reddit     & 672447  & 10984 & 1369       & 9578    & 37        & 0.004     & 78516  & 53589      & 17933   & 6994      & 0.281     \\
				lastfm     & 1293103 & 1980  & 227        & 1681    & 72        & 0.041     & 154993 & 106038     & 30883   & 18072     & 0.369     \\
				CanParl    & 74478   & 734   & 384        & 340     & 10        & 0.029     & 51331  & 38401      & 7432    & 5498      & 0.425     \\
				\bottomrule\bottomrule
				\end{tabular}
		}
	\end{table}

\section{Towards More Targeted Negative Sampling Strategies for Dynamic Link Prediction\label{sec:dlp}}
The Birth-Death diagrams introduced in Section \ref{sec:dynamic_graphs}, illustrate the partitioning of nodes and edges into distinct categories: Historical, Overlap, and Inductive. The comparison between a test event and a negative event involving an edge or nodes from any of these categories presents varying levels of difficulty for the task of discriminating the true event from the negative one. In this section, we operationalize these insights by formally defining Dynamic Link Prediction (DLP) and its connection to Negative Sampling (NS), before uncovering a taxonomy of the NS strategies targeting different aspects of DLP performance.

\subsection{Background on Dynamic Link Prediction}
Having thoroughly analyzed the nodes and edges in a CTDG, we now formalize the task of \emph{Dynamic Link Prediction (DLP)}. 

\begin{definition}
	The \textbf{DLP problem} is the task of distinguishing positive (true) interactions $(u,v,t)$ from interactions $(u',v',t)$ that do \emph{not} occur the same time $t$.
\end{definition}

For example, the task at hand can be to predict which two people $u$ and $v$, at the present time $t$, are most likely to interact in a social network.

\ptitle{Algorithms for DLP.} In practice, DLP algorithms are required to output a \emph{score} $s(u,v,t|\Hcal_t)$ that expresses the likelihood of the event $(u,v,t)$ given the past history of events $\Hcal_t$ up to time $t$ (as any future events would be unavailable at that time). When discussing DLP algorithms, it will then be helpful to also define $\Ucal_t$ as shorthand notation for the set of nodes interacting up to time $t$ and $\Ecal_t$ for the set of edges.
We note that parametric DLP algorithm (e.g. neural networks) will typically output a score $s(u,v,t|\Hcal_t)$ that is a function of the past history $\Hcal_t$ and of the parameters of the model. Thus, to evaluate the score a given test event $(u,v,t)$, DLP algorithms are given access to the history up to time $t$, but are not allowed to update their parameters based on these test events.

\ptitle{Negative Sampling for Evaluation.}
As already discussed, the goal of DLP algorithms is to accurately score the 
events $(u,v,t)$, conditioned on the past $\Hcal_t$ at time $t$. Ideally, a perfect model would score any \emph{positive} event $(u,v,t)$, i.e. an event that actually occurs in the history $\Hcal$, higher than any \emph{negative} event $(u',v',t)$ such that $(u',v')\neq (u,v)$, i.e. an event that could have occurred (but did not) at time $t$. The default strategy would thus fetch all possible edges and calculate their scores jointly with the positive at time $t$.
However, computing the scores for all possible edges scales quadratically with the number of nodes. Even for reasonably sized networks with a few thousand nodes, this renders the exhaustive comparison intractable. Consequently, as is the case in static link prediction, it is common to strongly subsample the set of possible negatives. We now formally define this crucial step of the evaluation.

\begin{definition}\label{def:neg_sampling}
	A \textbf{Negative Sampling} strategy is a mapping that takes as input a positive event $(u,v,t)\in\Hcal$ and returns a set of $K$ associated negative events $\{(u^{(k)},v^{(k)},t)\}_{k=1,\ldots,K}$ occurring at the same timestamp $t$.
\end{definition}

\ptitle{Problems with naive Negative Sampling strategies.}
A straightforward NS strategy is to swap the source $u$ and/or destination node $v$ of the positive event $(u,v,t)$ by other nodes $u' \in \mathcal{U}$ and/or $v' \in \mathcal{U}$ uniformly at random. Though common, this strategy is naive. The vast majority of possible negatives at time $t$ tends to be unrealistic for various intuitive reasons. For instance, many edges never occur at all in the graph, and many nodes only interact long before or after time $t$ (i.e. are probably \emph{inactive} at this time). 
Such unrealistic NS strategy may give an unbiased estimate of the exhaustive performance (obtained through comparison of the positive with all the possible edges). However, including trivial negatives into the comparison will steer the accuracy to a value close to 1, rendering such accuracy rather uninformative. To get accuracy estimates which align better with the actual task at hand, it is therefore important to only consider more challenging negatives. In what follows we will investigate how to do so.

\subsection{A Taxonomy of Negative Samples}
\label{sec:taxonomy}
Given that unrealistic NS leads to uninformative evaluation of DLP, how might we generate more useful negative samples? 

Clearly, this depends on the task at hand. For example, in the Flights dataset, the goal is to predict which destination a plane in a given origin airport will depart to at a given time. In this case, most edges will have been observed already, and it is more interesting to evaluate whether our model can distinguish between the actual origin and destination of the flight and an origin-destination pair that has been previously observed. Therefore, a realistic negative sample could be generated by replacing the actual edges with previously observed edges.

Similarly, in the case of social networks such as e-mail datasets, the goal may be predicting the receiver of a message emitted by a given sender at a certain time. In this case, one may want to specifically sample a negative edge $(u',v',t)$ such that the destination node $v'$ interacts in both the train and test set. The intuition for such a choice is that models will tend to naturally assign a higher score to events whose nodes have interacted in the train set. Moreover, the fact that these nodes are also present in the test set indicate that they may still be active at the time of the positive interaction, thus making the negative interaction a reasonable candidate.

Here we aim at contributing NS strategies which can be applied to any dataset or task.
Instead of focusing on specific applications, we present general taxonomy of potential edges that can be used as negative samples. DLP performance can then be appropriately scrutinized for different types of negatives in \emph{any} application. Considering that machine learning evaluation typically distinguishes performance on the train set from performance on the test set, we make heavy use of our definition of temporal categories in Sec.~\ref{sec:bd}.

Assume any NS strategy starts from a positive event $(u, v, t)$ and then `corrupts' it into (realistic) negatives $\{(u^{(k)},v^{(k)},t)\}_{k=1,\ldots,K}$. As discussed previously, the timestamp $t$ is left unchanged, as this is the time at which the prediction is assumed to be made. We can then either corrupt one of the nodes in the event, (which we call negative \emph{node} sampling) or both (which we call negative \emph{edge} sampling). For both, different strategies can be discerned, as described in the following two definitions:

\begin{definition}[Negative Node Sampling]
	\textbf{Negative Node Sampling} (Node-NS)  takes a positive event $(u, v, t)$ and replaces the source node $u$ by $u' \in \Ucal$ \emph{or} the destination node $v$ by $v' \in \Ucal$. In both cases, the other node is left the same.

	For a given cutoff time $t_{\text{split}}$, we distinguish six types of negative node samples:

	\begin{description}
		\item[Historical Source (HS)] $$d_\Hcal^{u'} < t_{\text{split}}$$
		\item[Overlap Source (OS)] $$b_\Hcal^{u'} < t_{\text{split}} \land d_\Hcal^{u'} \geq t_{\text{split}}$$
		\item[Inductive Source (IS)] $$b_\Hcal^{u'} \geq t_{\text{split}}$$ 
		\item[Historical Destination (HD)] $$d_\Hcal^{v'} < t_{\text{split}}$$
		\item[Overlap Destination (OD)] $$b_\Hcal^{v'} < t_{\text{split}} \land d_\Hcal^{v'} \geq t_{\text{split}}$$
		\item[Inductive Destination (ID)] $$b_\Hcal^{v'} \geq t_{\text{split}}$$
	\end{description}

	\begin{remark}
		In undirected graphs, source and destination sampling are equivalent.
	\end{remark}

\end{definition}

\begin{definition}[Negative Edge Sampling]
	 \textbf{Negative Edge Sampling}  (Edge-NS) takes a positive event $(u, v, t)$ and replaces the edge $(u, v)$ by another edge $(u', v') \in \Ucal \times \Ucal$.

	For a given cutoff time $t_{\text{split}}$, we distinguish three types of negative edge samples:
	\begin{description}
		\item[Historical Edge (HE)] $$d_\Hcal^{(u', v')} < t_{\text{split}}$$
		\item[Overlap Edge (OE)] $$b_\Hcal^{(u', v')} < t_{\text{split}} \land d_\Hcal^{(u', v')} \geq t_{\text{split}}$$
		\item[Inductive Edge (IE)] $$b_\Hcal^{(u', v')} \geq t_{\text{split}}$$
	\end{description}
\end{definition}

These strategies lend themselves to a straightforward visualization interpretation. For instance, on the Birth-Death diagrams (see for instance Fig. \ref{fig:bd_examples_edges}), HE, OE and IE correspond to sampling the negative edge respectively from the set of blue, orange or green points.

In the rest of this work, we will conduct experiments using the strategies HE, OE, and IE. These strategies enable us to compare the score of a positive event (an action that did happen) with the score of an event that occurred in the dataset but at a different time.
Furthermore, we will closely investigate the HD, OD, and ID strategies. These are particularly relevant for evaluating many DLP tasks (e.g. recommendation) where we compare the score of an interaction from a given source (such as a user) to the true destination (an item or another user) with the score of a negative destination that is randomly sampled. The HD, OD, and ID strategies help us understand the impact of choosing a negative destination from different time intervals. This choice can significantly affect the results of an evaluation.

\section{Experimental Setup}\label{sec:experiments}

In the remaining of the paper, we conduct extensive experiments in order to validate our evaluation tools, and answer two main questions.

\subsection{Datasets}

We selected 5 datasets of various sizes from a recent benchmark \cite{poursafaeiBetterEvaluationDynamic2023}. \textbf{Wikipedia} is a dataset of edits of Wikipedia pages recorded over one month. \textbf{Mooc} is a dataset of online behavior of students interacting with content (items) on a Mooc platform. \textbf{LastFM} is a dataset of interaction between users and songs listened by users. \textbf{UCI} is a communication network of university students exchanging over a social network. \textbf{Enron} is a dataset of emails between employees of a company, over a period of three year.

\subsection{Methods}
\label{sec:methods}
In our experiments, we consider 4 different DLP algorithms.
The first two methods are simple parameter-free heuristics, which work by memorizing either the \emph{nodes} or the \emph{edges} observed in past events. We describe them here, and detail the type of error that these are susceptible to commit:

\begin{itemize}
	\item \textbf{Preferential Attachment (PA)} assigns a score of $1$ to an event $(u,v,t)$ if and only if both the nodes have been observed in the past $\Hcal_t$:
	      \begin{equation}
		      s(u,v,t|\Hcal_t) = \ind{u\in \Ucal_t \wedge v \in \Ucal_t}.
	      \end{equation}
	      This method issues false negatives when the true event $(u,v,t)$ is such that either of $u$ or $v$ was never observed prior to $t$, and False Positive when the negative event $(u',v',t)$ is such that both $u'$ and $v'$ were involved in a past event.
	\item \textbf{EdgeBank} \cite{poursafaeiBetterEvaluationDynamic2023} assigns a score of $1$ to an event $(u,v,t)$ if and only if the edge $(u,v)$ has been observed in the past $\Hcal_t$:
	      \begin{equation}
		      s(u,v,t|\Hcal_t) = \ind{(u,v) \in \Ecal_t}.
	      \end{equation}
	      This method yields False Negatives whenever the true event $(u,v,t)$ is such that the edge $(u,v)$ was never involved in any event up to time $t$. It produces False Positives whenever the negative event $(u',v',t)$ is such that $(u',v')$ was involved in an event before $t$.
\end{itemize}

While simple, these baselines are helpful reference points to assess the performance of more sophisticated methods, as they relate directly to the Birth-Death diagrams presented in Sec. \ref{sec:bd_diagram}.
Indeed, one can think of Preferential Attachment and EdgeBank as follows. If we draw a horizontal line at the y coordinate corresponding to the current time $t$, PA will assign a score of 1 to any event $(u,v,t)$ such that both $u$ and $v$ are below the line, i.e. have a birth time prior to $t$: $b^u_{\Hcal}<t$ and $b^v_{\Hcal}<t$.
Similarly, EdgeBank will output a score of 1 to any event such that the \emph{edge} $(u,v)$ is represented by a point below the horizontal line with y coordinate equal to $t$ (i.e. $b^{(u,v)}_{\Hcal}<t$). On top of these methods, we consider two memory-based dynamic graph representation learning methods, which we briefly introduce here.


\begin{itemize}
	\item \textbf{TGN-attn} \cite{huangTemporalGraphBenchmark2023} is a DLP algorithm composed of two main modules. A memory module is responsible for maintaining a node-level memory state, encoding the past events at the node level. Using an attention module, the memory states of different nodes are then combined and subsequently used to calculate the probability of the event $(u, v)$.
	We use 
	\item \textbf{DyRep}\cite{trivediDyRepLearningRepresentations2018} has a similar architecture, but specifically uses an attention mechanism on the destination node in order to update the memory given an incoming event.
\end{itemize}

\begin{remark}
	In terms of implementation, we use a common approximation to the memory-based methods discussed above.
	Instead of updating the memory state (set of observed nodes or edges for PA/EdgeBank, node-memory states for TGN and Dyrep) at every newly observed events, the interactions are consumed by mini batches of 200 events.
	For each batch, the models successively compute a prediction score  for the events in this batch and the associated negative samples, and then update their memory by ingesting the events in that batch.
\end{remark}

\subsection{Metrics}

\ptitle{Binary Classification.}
In a first experiment, we view DLP as a binary classification task. For each positive event, we draw a negative event at random, following a specific NS strategy.
We thus obtain as a result a list of labeled events, where the label indicates whether it is an event that actually occurred or whether it is a negative sample. Considering a given threshold on the prediction scores, a \textbf{confusion matrix} such as the one in Table. \ref{tab:confusion_matrix} may then be constructed, measuring the amount of Positive/Negative events that get scored higher (positive prediction) or lower (negative prediction) than the threshold.
By varying the threshold, we can then draw the corresponding Receiver Operating Characteristic (ROC), and compute the associated \textbf{Area Under the Curve (AUC)}.
Note that this is typically done per batch of events, as the prediction scores may not be comparable across batches corresponding to different time spans. The reported AUC is the \emph{average} of the AUCs obtained on the different batches. 

\begin{table}[h]
	\caption{Confusion Matrix}
	\begin{center}
		\begin{tabular}{
			>{\centering\arraybackslash}m{0.2\linewidth}
			>{\centering\arraybackslash}m{0.2\linewidth}
			>{\centering\arraybackslash}m{0.2\linewidth}
			}
			\toprule
			\textbf{Predicted/Actual}   & \textbf{Actual Positive} & \textbf{Actual Negative} \\
			\midrule
			\textbf{Predicted Positive} & True Positive (TP)       & False Positive (FP)      \\
			\textbf{Predicted Negative} & False Negative (FN)      & True Negative (TN)       \\
			\bottomrule
		\end{tabular}
		\label{tab:confusion_matrix}
	\end{center}
\end{table}

\ptitle{Ranking.}
In order to assess which negative edge type is more likely to deceive the model at a given time, in a second experiment we consider ranking as a measure of performance. More precisely, we suppose that for each positive event, we assign it a certain number $K$ of negative events, obtained through various NS strategies. Then we calculate for each of these entries the rank of the positive but also of the associated negatives. Thus, for each event $(u,v,t)$ in the test data, we have a list of ranks corresponding to the positive event, and to different NS strategies, say NS1, NS2, etc. 

In order to get a regularly sampled time series, we then partition the time interval $[0,T]$ time into $B=50$ bins $I_1,...,I_{B}$ of equal size. For each interval we calculate the Mean Average Rank (MAR) of all the positives within that interval and do the same for each NS strategy. As a result, we get for each event type (positive, NS strategy 1, NS strategy 2 ...) a time series of the ranks of the associated events in the different intervals.

For instance, suppose that the first interval $I_1$ contains 4 positive events, to each of which we adjoin 2 negative events (coming from 2 different NS strategy NS1 and NS2). 
Now using the DLP algorithm, we obtain scores for each of the positives and the negatives. Suppose that, as result the ranks of the 4 positive events are 1, 3, 2, 1, the ranks of the negative obtained using NS1 are 2,1,3,2, and the ranks for NS2 are 3,2,1,3. 
Then for this interval, the MAR for the positive, NS1 and NS2 events are $\frac{7}{4}$, 2, and  $\frac{9}{4}$ respectively.

\section{Results}
In this section we discuss our experimental results, with the aim of answering two research questions. The goal is to assess the impact of the different NS strategies at the aggregate level (subsection \ref{sec:results_aggregate}) and over time (subsection \ref{sec:results_time}), 
\label{sec:results}
\subsection{How do performances of DLP algorithms vary over different NS strategies?}
\label{sec:results_aggregate}

\begin{figure}[t]
	\centering
		\begin{subfigure}[h]{\linewidth}
			\includegraphics[width=\linewidth]{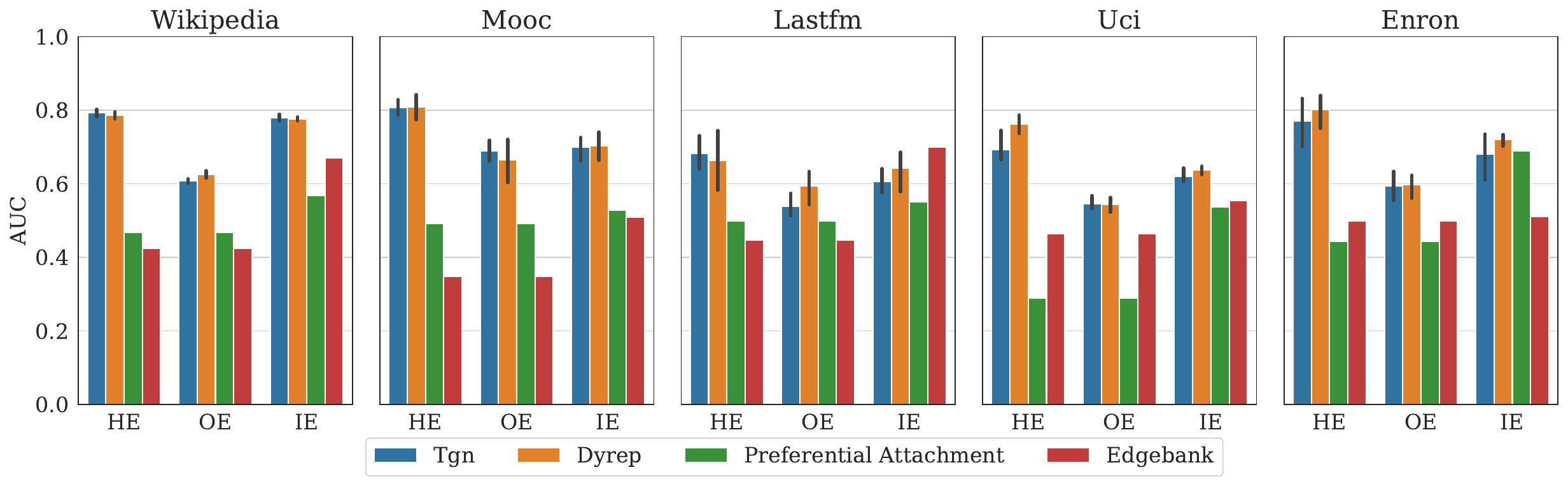}
			\caption{Test AUC results employing various Edge Negative Sampling strategies: Historical Edge (HE), Overlap Edge (OE), and Inductive Edge (IE). These strategies involve sampling negative edges (u',v') from sets corresponding to edges exclusively present during training, those present during both training and testing, and those exclusively present during testing, respectively.
			}
			\label{fig:one_vs_few_edge_ns}
		\end{subfigure}
		\hspace{20pt} 
		\begin{subfigure}[h]{\linewidth}
			\includegraphics[width=\linewidth]{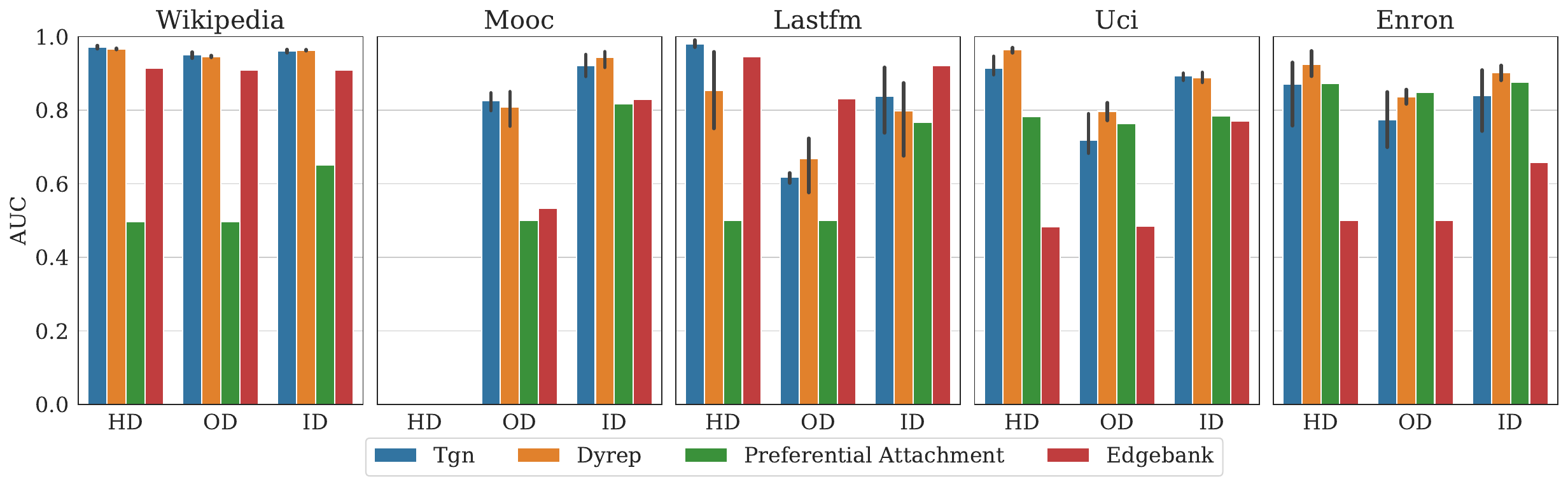}
			\caption{Test AUC results employing various Destination Negative Sampling strategies: Historical Destination (HD), Overlap Destination (OD), and Inductive Destination (ID). These strategies swap the destination node with one present during training, both training and testing, or exclusively during testing.}
			\label{fig:one_vs_few_dst_ns}
		\end{subfigure}
	\caption{Test AUC results obtained by comparing the scores of the positive events with the scores of the negative events, sampled using specific strategies. For Dyrep and TGN, we retrained the models with 5 different seeds and report the mean and the standard deviation of the resulting AUCs. }
	\label{fig:one_vs_few_aucs}
\end{figure}

In Figure \ref{fig:one_vs_few_edge_ns} we report the AUC scores against three Edge-NS: Historical Edge, Overlap Edge, and Inductive Edge Negative Sampling strategies. Similarly, in Fig. \ref{fig:one_vs_few_dst_ns} we report AUCs for three Destination-NS: Historical Destination, Overlap Destination, and Inductive Destination, as defined in Section \ref{sec:taxonomy}.

In general, the AUC scores are higher when swapping only the destination node, as in Fig \ref{fig:one_vs_few_dst_ns}. This makes sense when thinking that swapping the destination with a random nodes may lead to edges that never happened overall, and thus to negative events that are very unlikely.
The Overlap Edge and Overlap Destination seem to lead to the lower scores in general. Indeed, these strategies yield events that hit a good trade-off between having enough memory about the associated nodes/edges to yield a high score, and being sufficiently novel so that the models don't know yet how to discriminate them from the positive event. In contrast, for historical edges and destination, the involved nodes and edges become relatively obsolete after a certain time, an effect which seems to be picked up by the models.

Starting with the baselines, we note that EdgeBank and Preferential Attachment show very similar performances against Historical and Overlap Edges, with EdgeBank performing worse than random (AUC<0.5). This can be explained by reminding that, at test time, all the Historical Edge and Overlap Edge will have a score of 1. In contrast, the true event may have a score of 1 or 0 depending on whether the associated edge or nodes were previously observed. In that context, the false positive rate is greater than 0 only if the decision threshold is at least 1 (if it is below 1, all the negatives are predicted negative). When the decision threshold reaches the value 1, the number of false positives jumps to the number of negative (all the negatives will be predicted positives). In particular, the false positive rate will increase faster than the true positive rate, resulting in an AUC score lower than 0.5.


These figures indicate that NS can be defined such that heuristic baselines outperform neural-network based methods. For instance, in the IE and ID settings, EdgeBank is better than TGN and Dyrep on the LastFM dataset. Moreover, on the Enron dataset, PA outperforms the three other methods in the OD setting, and slightly outperforms TGN in the ID and IE setting. This is counter-intuitive when remembering that PA doesn't retain any information about the edges active in the past, but only remembers which \emph{node} was active in the past. 

 In general, TGN seems to yield a higher score than Dyrep. This makes sense considering that Dyrep has been shown to be a special case of TGN. However, there are some exceptions, notably in the Overlap destination setting where Dyrep is better on LastFM, UCI and Enron.

\def\figwidth{0.9\linewidth}

\begin{figure}
	\begin{subfigure}[t]{0.8\linewidth}
		\centering
		\includegraphics[width=\linewidth, trim={1.5cm 1.5cm 0 0cm}]{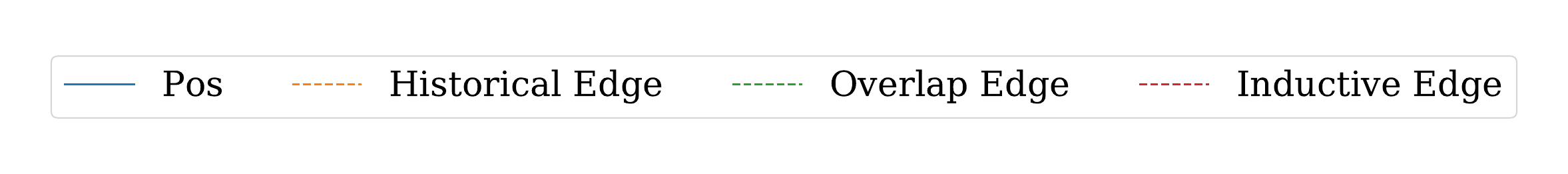}
	\end{subfigure}
		\begin{tabular}{p{0.5cm}l}
			PA       &
			\begin{subfigure}[t]{\linewidth}
				\centering
				\raisebox{-0.5\height}{\includegraphics[width=\figwidth,trim={0 0cm 0 0},clip]{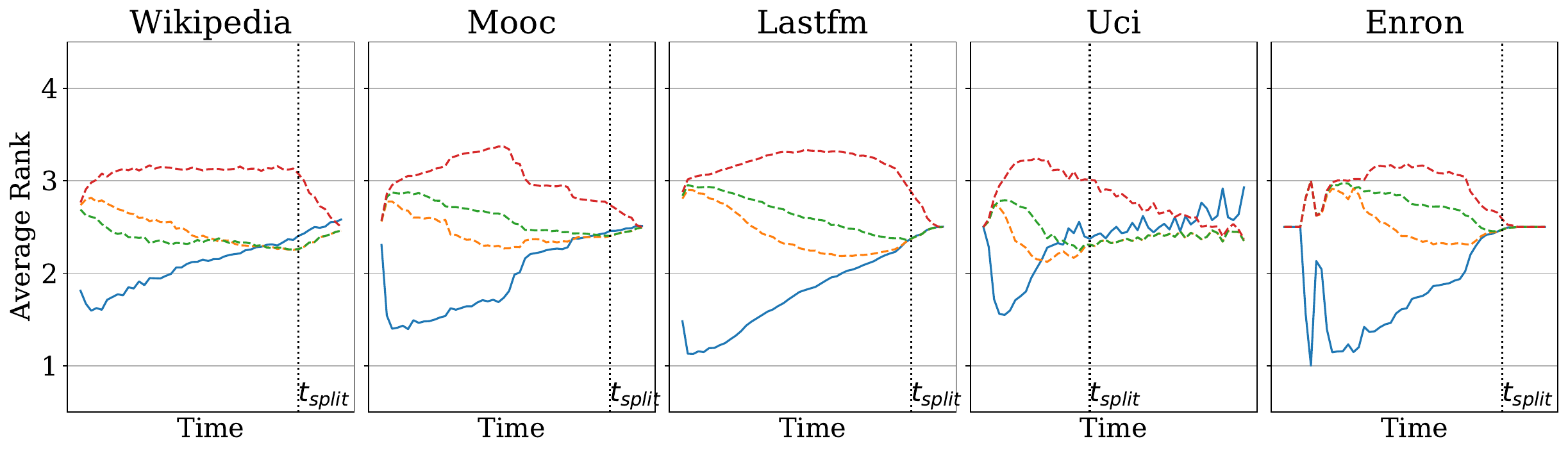}}
			\end{subfigure}  \\
			EdgeBank &
			\begin{subfigure}[t]{\linewidth}
				\centering
				\raisebox{-0.5\height}{\includegraphics[width=\figwidth,trim={0 0cm 0 1.1cm},clip]{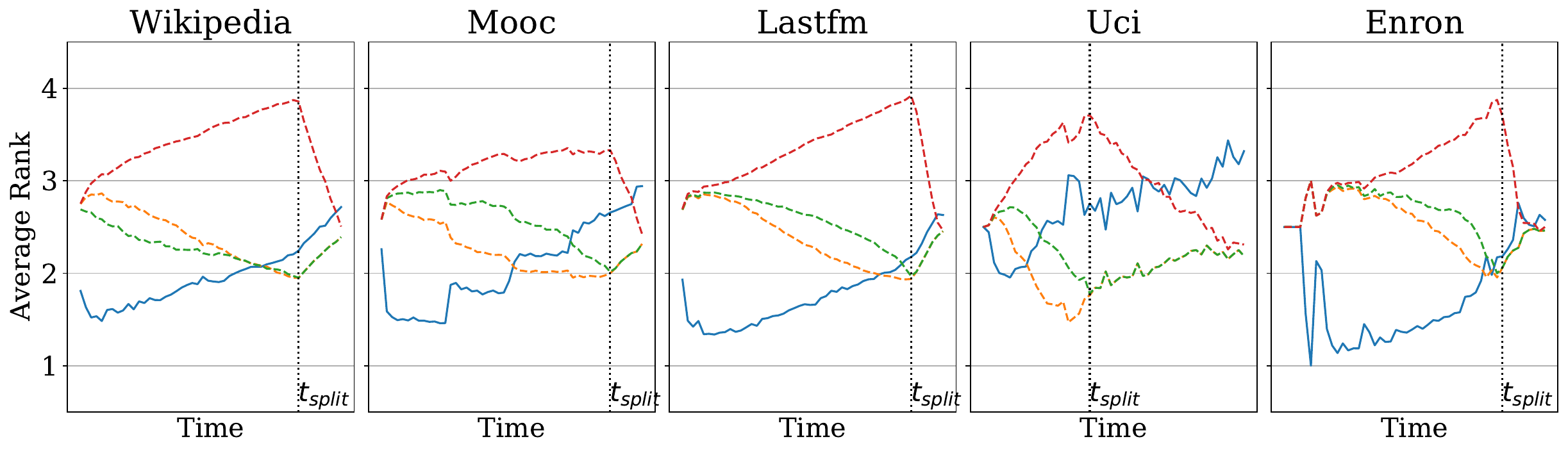}}
			\end{subfigure} \\
			Dyrep    &
			\begin{subfigure}[t]{\linewidth}
				\centering
				\raisebox{-0.5\height}{\includegraphics[width=\figwidth ,trim={0 0cm 0 1.1cm},clip]{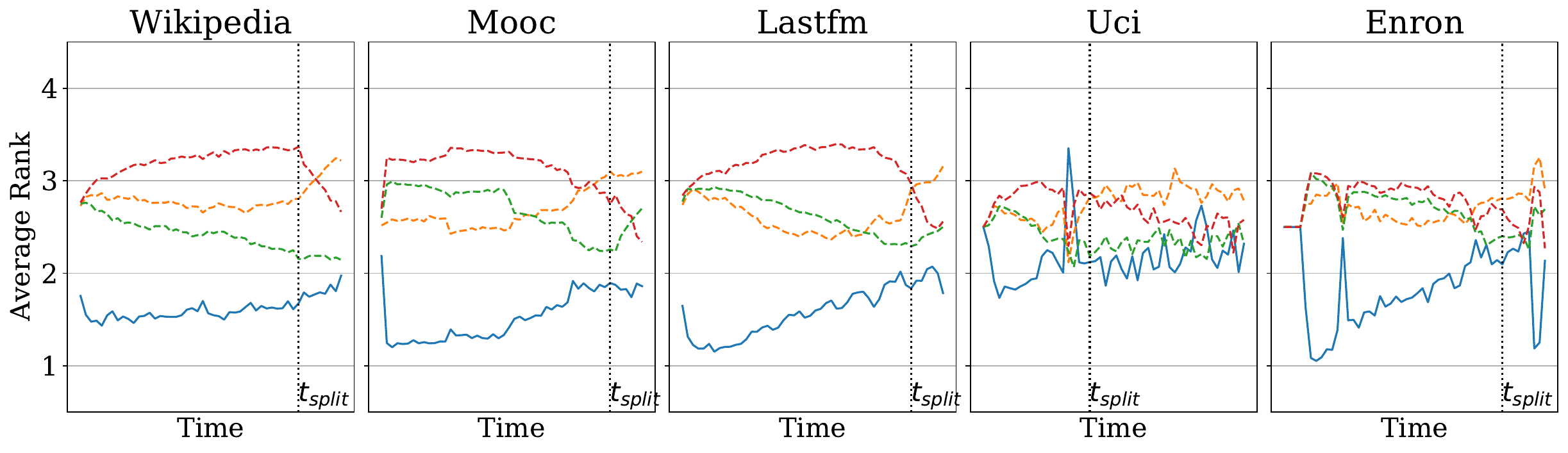}}
			\end{subfigure} \\
			TGN      &
			\begin{subfigure}[t]{\linewidth}
				\centering
				\raisebox{-0.5\height}{\includegraphics[width=\figwidth,trim={0 0cm 0 1.1cm},clip]{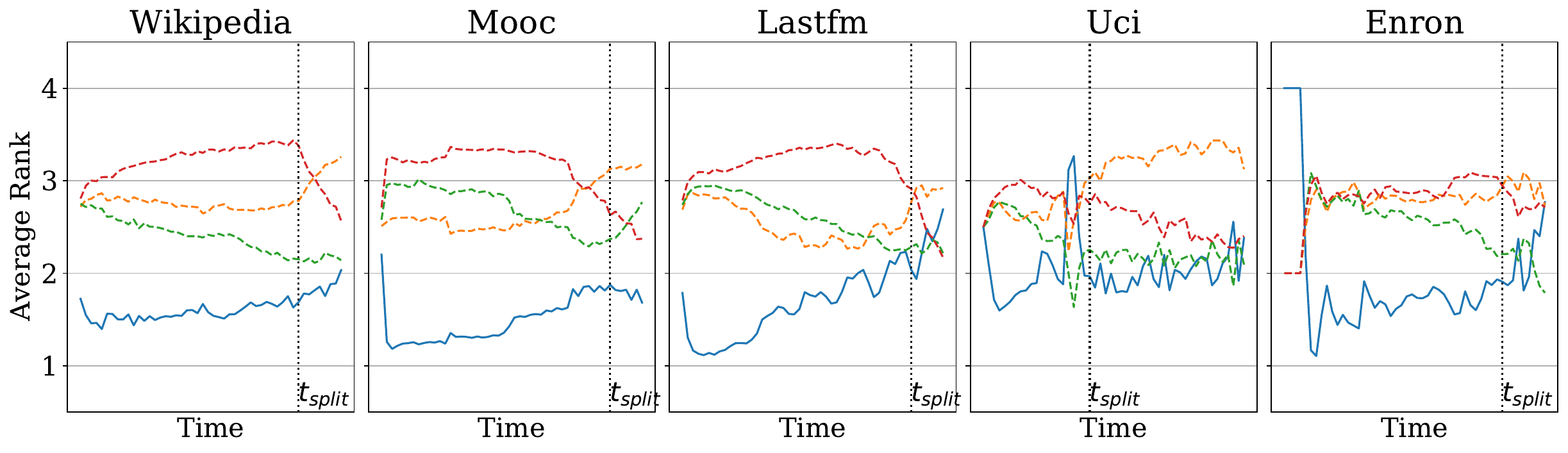}}
			\end{subfigure} \\
		\end{tabular}
	\caption{
		MAR of each method over time. The positive event "Pos" is ranked against negative events resulting from different Negative Edge Sampling strategies: Historical Edge, Overlap Edge, Inductive Edge, as defined in Section \ref{sec:taxonomy}.
	}
	\label{fig:long_one_vs_few_edge}
\end{figure}
\begin{figure}
	\begin{subfigure}[t]{0.8\linewidth}
		\centering
		\includegraphics[width=\linewidth, trim={1.5cm 1.5cm 0 0cm}]{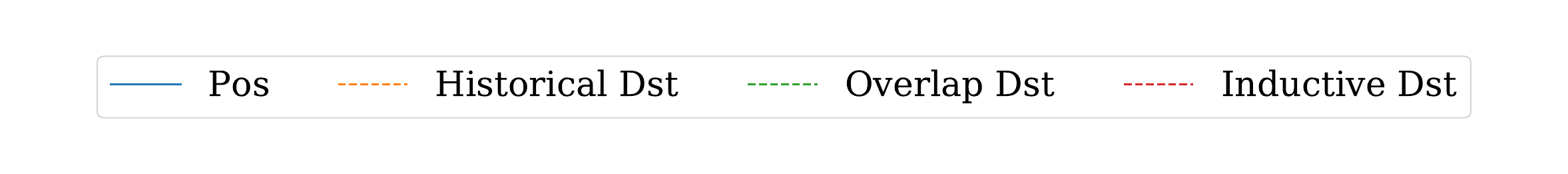}
	\end{subfigure}
		\begin{tabular}{p{0.5cm}l}
			PA       &
			\begin{subfigure}[t]{\linewidth}
				\centering
				\raisebox{-0.5\height}{\includegraphics[width=\figwidth,trim={0 0cm 0 0},clip]{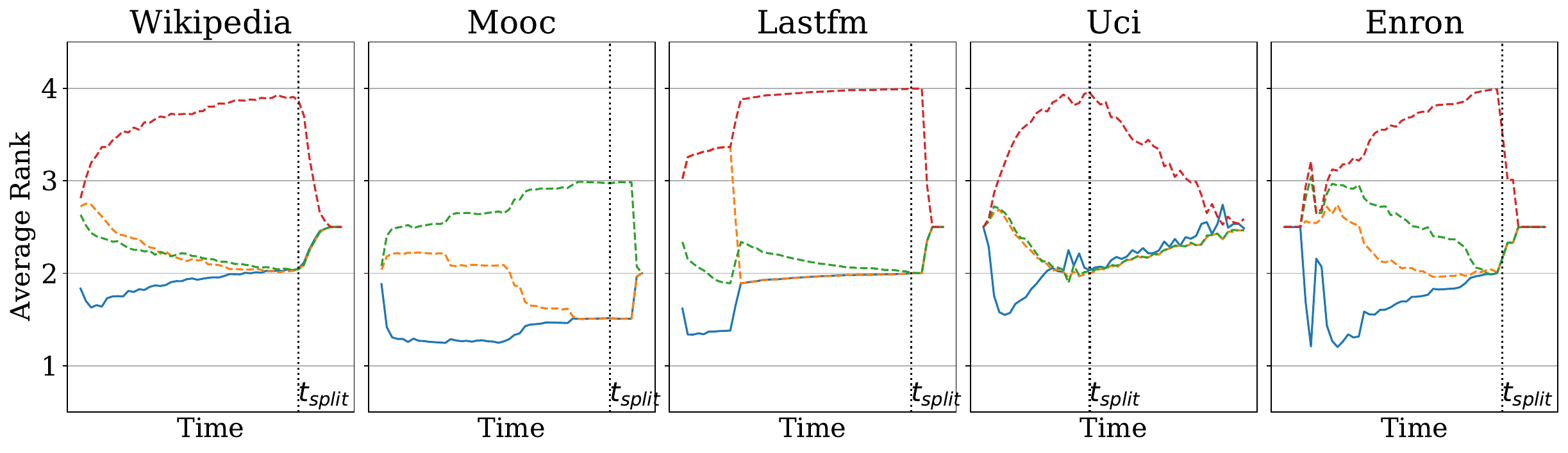}}
			\end{subfigure} \\
			EdgeBank &
			\begin{subfigure}[t]{\linewidth}
				\centering
				\raisebox{-0.5\height}{\includegraphics[width=\figwidth,trim={0 0cm 0 1.1cm},clip]{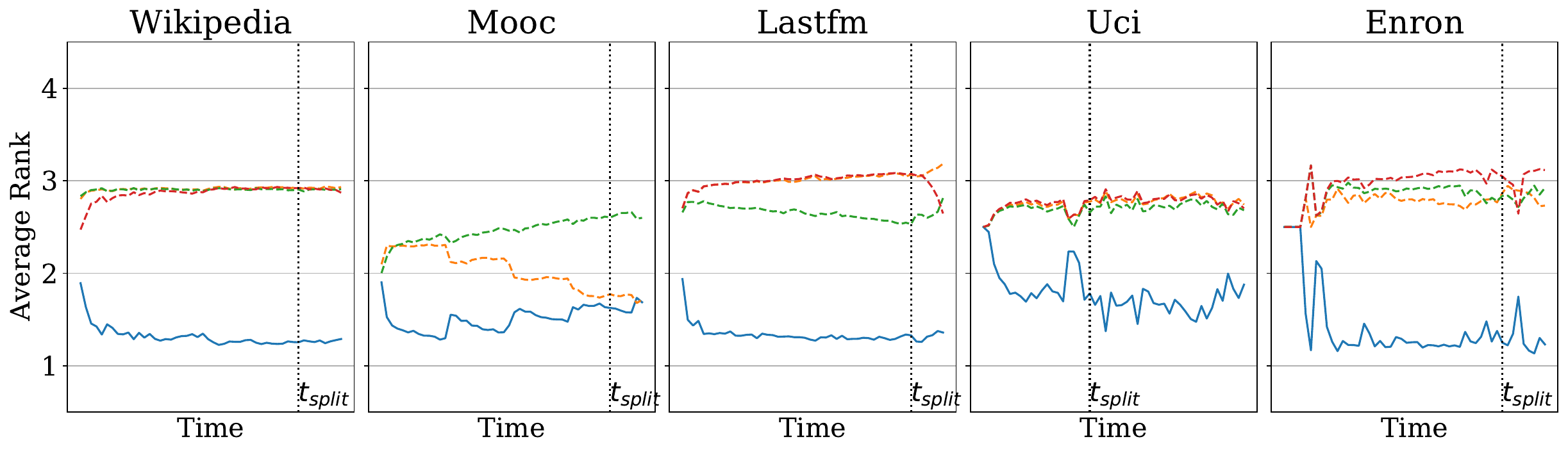}}
			\end{subfigure} \\
			Dyrep    &
			\begin{subfigure}[t]{\linewidth}
				\centering
				\raisebox{-0.5\height}{\includegraphics[width=\figwidth ,trim={0 0cm 0 1.1cm},clip]{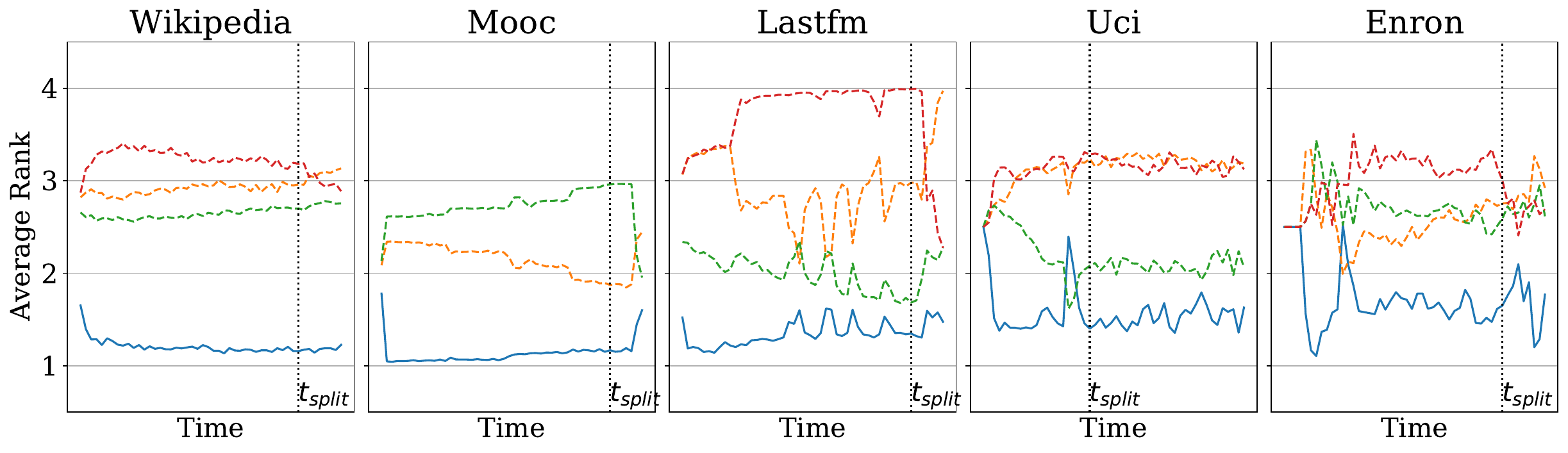}}
			\end{subfigure} \\
			TGN      &
			\begin{subfigure}[t]{\linewidth}
				\centering
				\raisebox{-0.5\height}{\includegraphics[width=\figwidth,trim={0 0cm 0 1.1cm},clip]{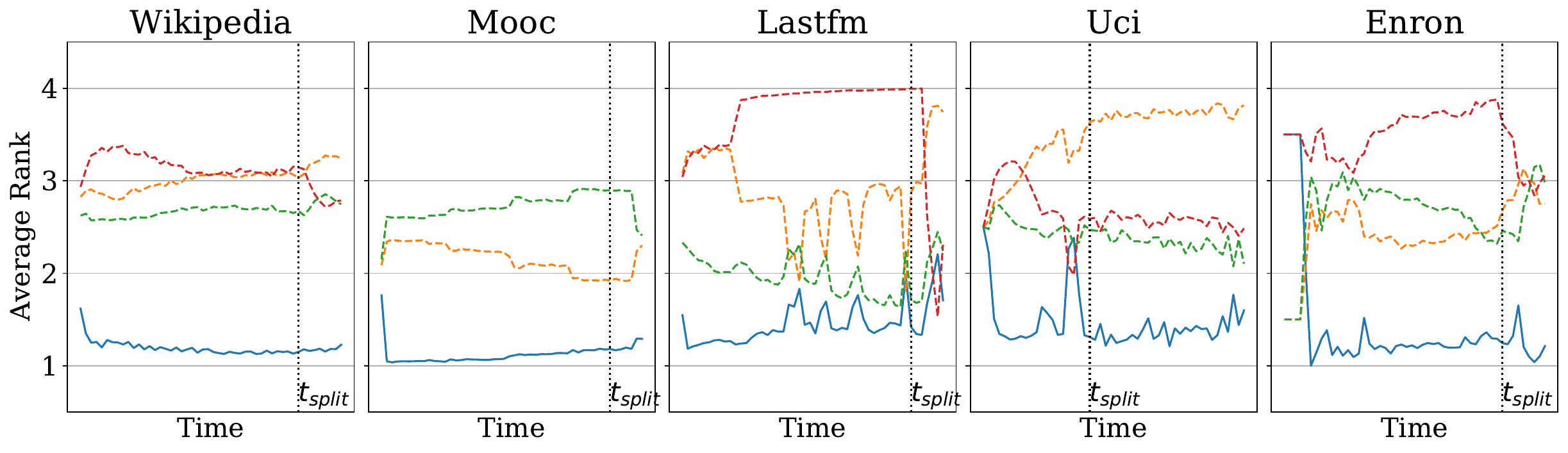}}
			\end{subfigure} \\
		\end{tabular}
	\caption{
		MAR of each method over time. The positive event "Pos" is ranked against negative events resulting from different Negative Destination Sampling strategies: Historical Destination, Overlap Destination, Inductive Destination, as defined in Section \ref{sec:taxonomy}.
	}
	\label{fig:long_one_vs_few_dst}
\end{figure}

\subsection{Which competing Negative Sampling strategy misleads the model more over time?}
\label{sec:results_time}

In the previous experiment, we observed that using a different NS strategy can dramatically alter the prediction results. However, it is crucial to note that performance evolves over time in the Temporal Graph. 
In this section, we demonstrate how distinguishing different NS techniques can help provide more nuanced insight into the performance of each method over time.


On Figure \ref{fig:long_one_vs_few_edge} we observe the performance over time, against Edge-NS. For all methods and datasets, we note that the rank of the positive events increases over time while the rank of the other edges decreases. 
Indeed, as memory gets filled over time, more events will appear to be likely. In particular, the score of randomly sampled Historical and Overlap edges will go up in general as compared to the score of the positive event.

A general trend is that the rank of Inductive edge/destination NS (green and red lines respectively) increases on the training period, before dropping on the test set. This type of negative samples will eventually be the ones which will confuse the model during test time, as their rank becomes closer and closer to the rank of the positive event after $t_{split}$. In general, historical and Overlap items follow the opposite trend: they become more likely up to $t_{split}$, and their rank then starts increasing again either at $t_{split}$, or shortly before (notably for historical edges/destinations).

These plots make it clear that the \emph{type} of error made by the models changes over time. Indeed, while on the train set, Inductive edges are likely to be scored low (less likely) since they have not been observed yet, they become much more likely on the test set. During the test period even, for instance looking at EdgeBank on the UCI dataset in Fig. \ref{fig:long_one_vs_few_edge}, shortly after $t_{split}$, the model will tend to rank the positive events lower than the Overlap and historical edges. After some time, however, the Inductive edge negative samples will tend to be ranked similarly to the historical and Overlap. 

These visualizations give profound insight on the differences in performance between TGN and Dyrep. For instance on the LastFM dataset, TGN seems to be more consistent in ranking the Inductive Destination high compare to the other destination nodes. However, its overall performance on the Test set is not clearly better than Dyrep. On the other hand, in Fig. \ref{fig:long_one_vs_few_dst}, the performances on the UCI dataset indicate that, while TGN is clearly able to push historical destinations further than the Inductive destination, Dyrep tends to assign them similar rankings across the test set. 

On both Figures \ref{fig:long_one_vs_few_edge} and \ref{fig:long_one_vs_few_dst}, the examples on the UCI data allow to visualize the effect of seasonality on the performance. Indeed, for both methods, the rank of the positive event peaks just before the split time. This time corresponds roughly to the lower activity also observed in Figure \ref{fig:bd_uci}.

To conclude, the results shown in Fig. \ref{fig:one_vs_few_aucs}, \ref{fig:long_one_vs_few_dst} and \ref{fig:long_one_vs_few_edge} demonstrate that overly optimistic AUC may hide a more intricate reality, especially when taking simple baselines as reference point.
While more targeted NS strategies helps identify settings where heuristics outperform proposed heavier models, plotting the performance over time is important to get an idea of how a model's performance will react to domain-specific changes in the data over time.

\def\methods{
	dyrep,
	tgn,
	edgebank}
\def\datasetssmall{
	enron,
	lastfm,
	mooc,
	uci,
	wikipedia}

\clearpage
\section{Guidelines for Practitioners}
Before concluding this study, we propose a series of guidelines to practitioners of DLP, with the hope of further improving the evaluation of this task.

\begin{enumerate}
	\item One key take-away is that evaluating DLP algorithms requires exploring the performance along several dimensions of the data, at the node, edge and time-interval level. 
	\item To do so, as much as possible, a good practice for DLP algorithm is to enable saving the list of scores of each positive event and negative event. Indeed, these scores can be the starting point of in-depth evaluations such as the one conducted in this paper. As suggested in \cite{MARA2022evalne}, this type of standardized output format is critical in order to easily apply diverse evaluation methods, while minimizing evaluation error. 
	\item Birth-Death diagrams are important tools to understand the effect of time-based train-test splitting on the nodes and edges involved in the graph, and may help  hypothesizing the expected performance of a given method on a given dataset. 
	\item Simple baselines such as EdgeBank and Preferential Attachment Baselines are indispensable in order to understand whether the model learns anything non-trivial from the data.
	\item Finally, visualizing how the prediction performance evolves over time can be crucial in understanding the strengths and weaknesses of DLP algorithms.
\end{enumerate}

\section{Conclusion}


Recent academic efforts have been dedicated to standardizing Dynamic Link Prediction (DLP) as a machine learning task. The goal is to equip the task with its own evaluation pipelines, baseline methods, and benchmark datasets. However, as a consequence of the high-dimensionality of the data and its non-independent, non-identically distributed nature, deriving a single consistent model validation has encountered numerous challenges. 

In this paper, we have explored several key aspects of these challenges. On the one hand we have investigated the effect of time-based train test splitting on the set of nodes and edges through the novel Birth-Death diagrams, and discussed examples of these visualizations on datasets from diverse domains. Moreover, we have shown how to rely on the proposed Birth-Death diagrams to derive more challenging negative samples, based on the hypothesis that the error depends on the NS strategy. To illustrate the effect of these negative samples, we conducted an empirical assessment of the impact of different negative samples on performance was conducted. The relative performance of methods across datasets was compared over time by plotting the prediction ranks as a time series, revealing interesting insights into the failure modes of the different methods.

This work raised several open questions that can be explored in future work. First, our evaluation tools could be used to conduct a more exhaustive comparison of existing heuristics such as the ones introduced in \cite{liben-nowellLinkpredictionProblemSocial2007}, first with simple learning based approaches such as the ones discussed in Sec.~\ref{sec:related_work}, and then with more recent representation learning methods, with an emphasis on fair comparison.
In terms of visualization, the proposed Birth-Death diagrams could be leveraged to visualize higher order structures such as cliques or triangles, or any repeatedly occurring subgraph that can be uniquely identified.

\clearpage

\appendix

\section[\appendixname~\thesection]{Birth-death diagrams on User-Item graphs}
\label{appendix:user_item}

\def\figwidth{0.5\linewidth}
\begin{figure}
	\begin{subfigure}[t]{\figwidth}
		\centering
		\includegraphics[width=\linewidth]{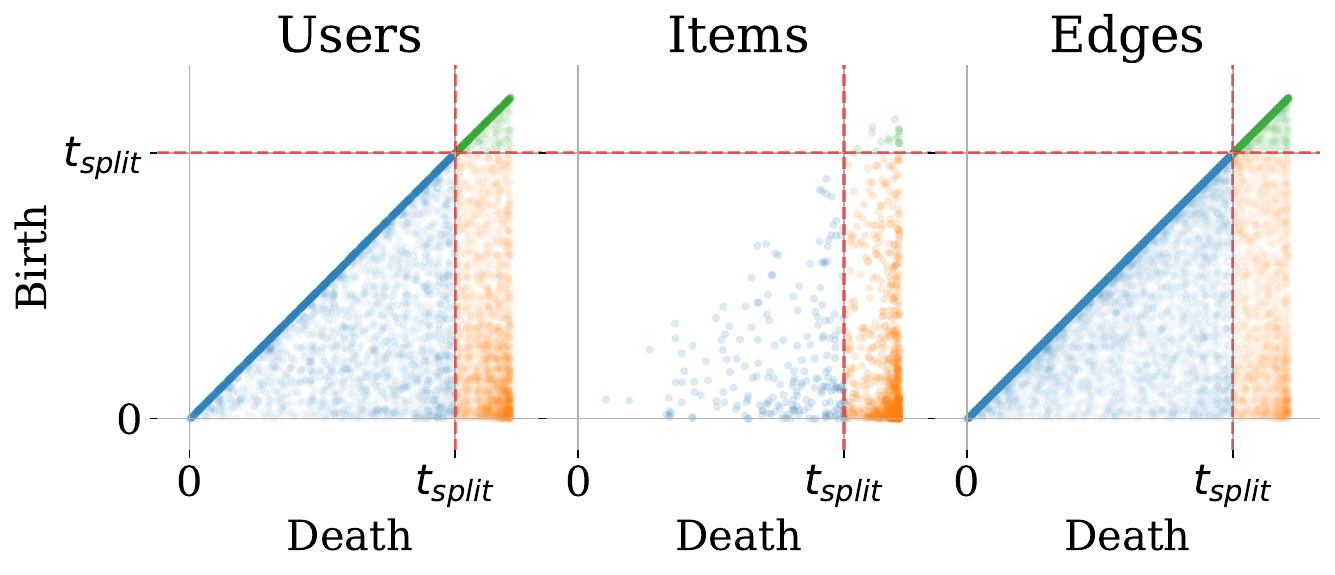}
		\caption{wikipedia }
	\end{subfigure}
    \begin{subfigure}[t]{\figwidth}
		\centering
		\includegraphics[width=\linewidth]{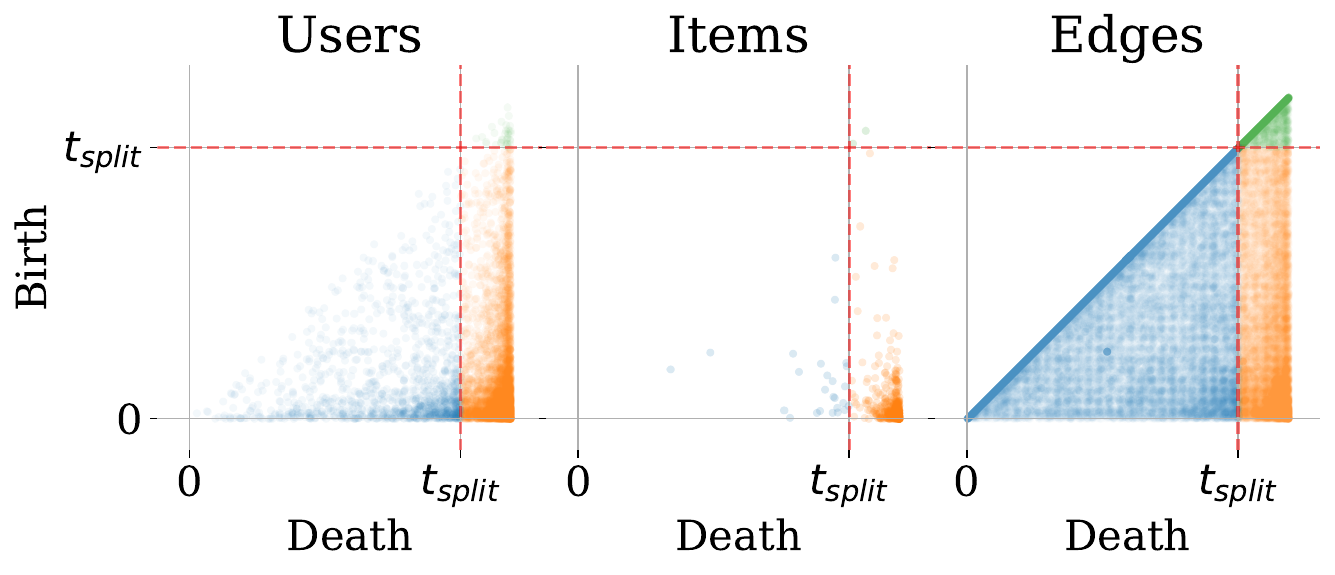}
		\caption{reddit }
	\end{subfigure}

    \begin{subfigure}[t]{\figwidth}
		\centering
		\includegraphics[width=\linewidth]{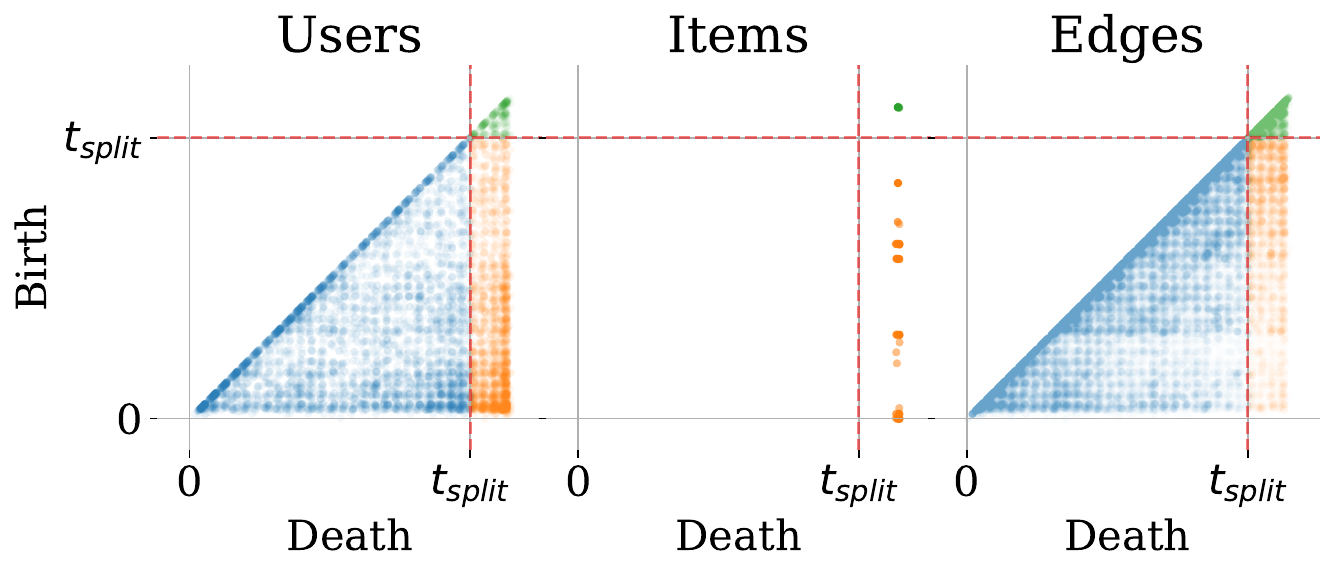}
		\caption{Mooc }
	\end{subfigure}
    \begin{subfigure}[t]{\figwidth}
		\centering
		\includegraphics[width=\linewidth]{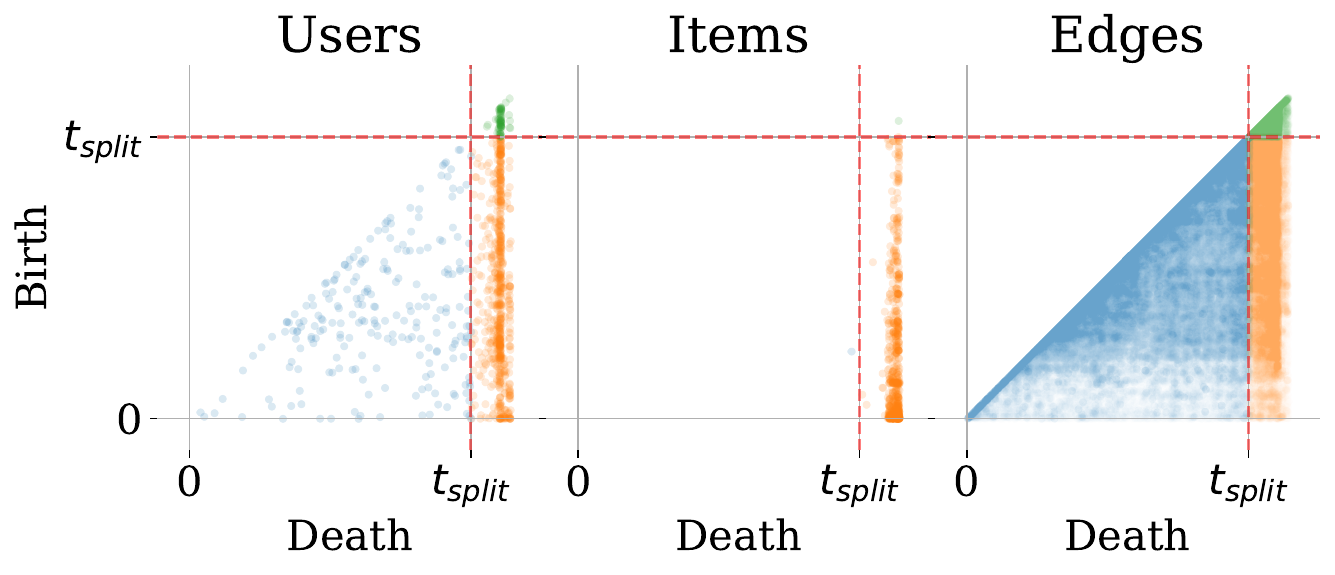}
		\caption{LastFM }
	\end{subfigure}
\caption{Birth-death diagrams on Bipartite User-Item graphs. }
\label{fig:bd_user_item}
\end{figure}

The dynamic graphs studied in this paper can be classified into three distinct types:

\begin{itemize}
    \item Unipartite Undirected. for instance, in face-to-face interaction dynamic graphs such as the HighSchool data, there is only one type of nodes (students), and an interaction between nodes $u$ and $v$ doesn't have a direction. Contact, SocialEvo, UNvote, USLegis are examples of such graphs.
    \item Unipartite directed: for example, in the Enron e-mail dataset, there is only one type of nodes (users involved in mail exchanges). However, the interactions (e-mails) have an orientation: a given user $u$ sends an e-mail \emph{to} user $v$. The datasets  UCI, UNtrade, Flights, and Enron are examples of such graphs.
    \item Bipartite: in this case there is a clear separation of the nodes into two node types. Typically, in \emph{User-Item} graphs such as wikipedia,LastFM, Mooc, reddit, users are always involved in interactions as senders, while items always receive an interaction (a click, a like, a subscription etc...).
\end{itemize}

As mentioned in the main body of the paper, Birth-Death diagrams allow visualizing when nodes and edges start and stop interacting for the first time in the history of events.
However, for Bipartite datasets, there is a clear separation between the nodes that will be involved in events as \emph{source} nodes and those that will appear as \emph{destination} nodes. 
For these specific datasets, plotting the birth and death time of users and items separately yields extra information. On Fig. \ref{fig:bd_user_item}, it can be observed for instance that in all dataset, most of the \emph{items} are overlap nodes, in the sense that they are observed in both the train and test set. In the Mooc dataset, all the items (courses that can be followed by students) are actually observed at least once during the test period, thus there are no historical destination nodes in that case. For the LastFM dataset, This is important since for the nodes in the lower right, prediction methods has the chance of accumulating a lot of information about them over time. This is in contrast with user nodes, where it can be seen  that users more commonly appear and disappear both during the train period.

\section{Model Architectures and Hyperparameters used in the experiments}
In the main paper we compared the performances TGN-Attn and Dyrep with heuristic baselines.
We used the implementation provided in the examples of the open-source TGB library 
\url{https://github.com/shenyangHuang/TGB/tree/main/examples/linkproppred/tgbl-wiki}.

The architecture of the TGN model is the following:
\begin{enumerate}
  \item The \emph{message function} is the identify function (same as in the original paper \citep{Rossi2020}).
  \item The \emph{aggregation function} is the last message aggregator (we keep for each node only the last message received from the batch).
  \item The \emph{memory updater} is a GRU.
  \item The \emph{embedding} module is Temporal Graph Attention Layer. Its purpose is to integrate both the network's connectivity information and temporal data, ensuring that embeddings remain current by combining the memory state with the most recent network information.
  \item The \emph{edge-level decoder} (i.e. Link Predictor) is a 2-layer MLP with 100 hidden units and ReLU activation.
\end{enumerate}

The Dyrep model implemented in the library is very similar to the TGN architecture, with two main differences:

\begin{itemize}
	\item The \emph{memory updater} is a simple RNN.
	\item The \emph{embedding} module is the identity: the memory is used directly for prediction. Note that this makes the model vulnerable to the memory staleness problem.
	\item However, the \emph{messages function} is calculated using a graph attention module on the destination node.
\end{itemize}

\ptitle{Hyperparameters.} As the goal of our study is not to maximize the performance but to compare it across Negative Sampling strategy, we used the default hyperparameters values for both TGN-Attn and Dyrep. Thus, the memory, time encoding, embedding dimensions are all set to 100. We use the Adam optimizer\cite{kingma2017adam} with a learning rate of $1e-4$ and a weight decay of $1e-4$. To prevent overfitting we use early stopping on the validation AUC: we stop the training if there has been no improvement in AUC of more than $1e-3$ in the last $20$ epochs.









\bibliographystyle{plainnat}
	
\bibliography{References.bib}

%



\end{document}